\documentclass[ba]{imsart}
\pubyear{2022}
\volume{TBA}
\issue{TBA}
\doi{10.1214/21-BA1303}
\arxiv{2104.09237}
\firstpage{1}
\lastpage{24}

\usepackage{amsthm}
\usepackage{amsmath}
\usepackage{natbib}
\usepackage[colorlinks,citecolor=blue,urlcolor=blue,filecolor=blue,backref=page]{hyperref}
\usepackage{amsfonts}
\usepackage{amssymb}
\usepackage{graphicx}
\usepackage{float}
\usepackage{cleveref}
\usepackage{hyperref}
\usepackage[flushmargin]{footmisc}
\usepackage{float}
\usepackage{xcolor}
\usepackage{bbm}
\usepackage{titlesec}
\usepackage{nameref}
\usepackage[toc,page]{appendix}
\usepackage{booktabs}
\usepackage{multirow}
\usepackage{multicol}
\usepackage{caption}
\usepackage{subcaption}

\graphicspath{{./Figures/}}

\startlocaldefs
\numberwithin{equation}{section}
\theoremstyle{plain}

\endlocaldefs

\DeclareMathOperator*{\argmax}{arg\,max}

\newcommand{\bs}{\boldsymbol}

\newcommand{\icol}[1]{% inline column vector
  \left(\begin{smallmatrix}#1\end{smallmatrix}\right)%
}

\DeclareUnicodeCharacter{2061}{}
\def\spacingset#1{\renewcommand{\baselinestretch}{#1}\small\normalsize} \spacingset{1}

\begin{document}

\begin{frontmatter}
\title{Inverse Bayesian Optimization: Learning Human Acquisition Functions in an Exploration vs Exploitation Search Task}
\runtitle{Learning Human Acquisition Functions}

\begin{aug}
\author{\fnms{Nathan} \snm{Sandholtz}\thanksref{addr1,t1,t2,m1}\ead[label=e1]{nsandholtz@stat.byu.edu}},
\author{\fnms{Yohsuke} \snm{Miyamoto}\thanksref{addr2,t3,m1,m2}\ead[label=e2]{second@somewhere.com}},
\author{\fnms{Luke} \snm{Bornn}\thanksref{addr3,t1,m2}%
\ead[label=e3]{third@somewhere.com}%
\ead[label=u1,url]{lukebornn.com}}
\and
\author{\fnms{Maurice A.} \snm{Smith}\thanksref{addr4,t1,m2}%
\ead[label=e4]{mas@deas.harvard.edu}%
\ead[label=u2,url]{a_url.com}}

\runauthor{N. Sandholtz et al.}

\address[addr1]{Department of Statistics, Brigham Young University
    \printead{e1} 
}
\address[addr2]{Waymo LLC
}
\address[addr3]{Zelus Analytics
}
\address[addr4]{School of Engineering and Applied Science, Harvard University
}

\end{aug}

\begin{abstract}

This paper introduces a probabilistic framework to estimate parameters of an acquisition function given observed human behavior that can be modeled as a collection of sample paths from a Bayesian optimization procedure.  The methodology involves defining a likelihood on observed human behavior from an optimization task, where the likelihood is parameterized by a Bayesian optimization subroutine governed by an unknown acquisition function. This structure enables us to make inference on a subject's acquisition function while allowing their behavior to deviate around the solution to the Bayesian optimization subroutine.   To test our methods, we designed a sequential optimization task which forced subjects to balance exploration and exploitation in search of an invisible target location.  Applying our proposed methods to the resulting data, we find that many subjects tend to exhibit exploration preferences beyond that of standard acquisition functions to capture.  Guided by the model discrepancies, we augment the candidate acquisition functions to yield a superior fit to the human behavior in this task.  

\end{abstract}

%\begin{keyword}[class=MSC]
%\kwd[Primary ]{60K35}
%\kwd{60K35}
%\kwd[; secondary ]{60K35}
%\end{keyword}

\begin{keyword}
\kwd{Bayesian optimization}
\kwd{directional statistics}
\kwd{exploration vs. exploitation}
\kwd{human cognition}
\kwd{inverse optimization}
\kwd{lab experiment}
\kwd{probabilistic models}
\end{keyword}

\end{frontmatter}

\section{Introduction}

The problem we study in this paper is a type of supervised learning problem: given observed human behavior from a sequential optimization task, we aim to create a model that accurately describes how they will behave on future rounds of the task.  However, unlike most supervised learning problems, we explicitly assume that the observed behavior in each round of the task represents an approximate solution to a latent optimization problem faced by the subject.  Not only do we want to incorporate this optimization subroutine into our model, but inferring the criteria governing the subject's optimization strategy is actually our primary interest.  This type of analysis can be termed an \textit{inverse optimization}.  

A notable gap within the inverse optimization literature is the lack of statistical treatment of its methodology.  Only very recently have statistical concepts such as consistency \citep{aswani2018inverse}, goodness of fit \citep{chan2019inverse}, and robustness \citep{ghobadi2018robust, esfahani2018data, shahmoradi2019quantile} been treated in methodological papers.  This is surprising because many optimization problems are assumed to be carried out imperfectly, thereby introducing noise to the data.  In such cases, inverse optimization becomes an inference problem well-suited to probabilistic modeling.   A key contribution of our work is the formulation of an inverse decision problem via a probabilistic model, which provides a principled framework to quantify both the variability with which the subject performs the optimization as well as the uncertainty around the parameter estimates governing the underlying optimization strategy.  

We illustrate our methods by analyzing human decision-making behavior from a sequential optimization task developed in collaboration with the Neuromotor Control Lab at Harvard University.  As will be further explained in Section \ref{sec:task_desc}, the task was specifically designed to present subjects with a conflict between exploration and exploitation on the second move of the task.  Our goal is to make inference on how the subjects strategically balance exploration and exploitation based on their behavior on this second move.

In order to make inference on the subjects' strategies, we require a model for their optimization procedure.  As will be detailed in Section \ref{sec:BO}, our model framework draws on Bayesian optimization, which is a sequential model-based approach to maximize (or minimize) an unknown objective function \citep{jones1998efficient, shahriari2015taking}.\footnote{Bayesian optimization is closely related to (or synonymous with, in some cases) sequential design of experiments, efficient global optimization, adaptive sampling, active learning, optimal search, and hyperparameter optimization.}  The core idea of Bayesian optimization is to characterize the uncertainty about an unknown objective function with a statistical model, termed a \textit{surrogate}, then to strategically synthesize the model uncertainty via an \textit{acquisition function} in order to determine promising new locations to sample.  We adopt this terminology in our paper, as the fundamental components of our optimization model follow those of Bayesian optimization.  

Within this context, our goal becomes to characterize and estimate subjects' acquisition functions (i.e. their exploration/exploitation preferences) based on their observed behavior.  The novelty of our method is that we define a likelihood on a subject's observed search paths parameterized by an underlying Bayesian optimization subroutine governed by an acquisition function.  This structure enables us to make inference on the subject's acquisition function while allowing their behavior to deviate around the solutions to the underlying optimization subroutine.  By using probabilistic models of this form, both the variability with which subjects optimize as well as the uncertainty around their estimated exploration/exploitation preferences can be quantified.  

We find that a wide range of acquisition functions are exhibited across the subjects in our study, but probability of improvement (PI) and upper confidence bound (UCB) functions tend to fit the majority of subjects best.  Unlike previous studies of this problem, we are able to provide credible intervals to characterize the uncertainty in the estimated acquisition functions.  Finally, there are many subjects for which none of the candidate acquisition functions we initially consider provide a satisfactory model of their behavior.  Guided by the model discrepancies for these subjects, we propose an augmentation to the acquisition functions to construct a superior model of human optimization behavior in this task.
 
\subsection{Related work and contributions}

A number of recent papers investigate the correspondence between ML algorithms and actual human behavior for various decision-making processes \citep{borji2013bayesian, wilson2015human, schulz2015assessing, wu2018generalization, plonsky2019predicting, gershman2019uncertainty, candelieri2020modelling}.  Within this context, \cite{borji2013bayesian} compare several ML techniques against observed human behavior in a set of sequential optimization tasks that require subjects to make choices along one continuous dimension, finding that Bayesian optimization algorithms best approximate the human behavior.  Recent papers expand their work to tasks that require discrete choices \citep{wu2018generalization} and tasks where the decision space is continuous in two dimensions \citep{candelieri2020modelling}.  These studies essentially explore the inverse problem of Bayesian optimization.  Our research builds on these studies by introducing a probabilistic framework for the inverse problem, which we introduce in Section \ref{sec:IBO}.  

More broadly, our work contributes to the literature on inverse optimization \citep{ahuja2001inverse} and related problems such as inverse reinforcement learning \citep{ng2000algorithms} and inverse decision theory \citep{swartz2006inverse}.  \cite{li2020inverse} is particularly relevant to our project, which uses inverse optimization to learn a convex function representative of a subject's risk preferences.  The estimand in our project---the acquisition function of a Bayesian optimization---is analogous to the risk preferences estimated in Li's paper. 

\subsection{Outline}

The rest of this paper is outlined as follows.  In Section \ref{sec:task_desc}, we describe the search task and explain how it forces subjects to make an exploration vs exploitation trade-off on the second move of the task.  In Section \ref{sec:BO}, we frame the search task as a Bayesian optimization procedure and show how the forward optimization proceeds under three acquisition function classes.   Section \ref{sec:IBO} introduces the inverse problem and illustrates our proposed solution framework based on the second move of the search task.  We also augment the acquisition functions to enable a more accurate fit to the human tendencies we observe.  In Section \ref{sec:human_tendencies} we present our results and explore relationships between the estimated acquisition functions and subjects' corresponding performance in the task.  Section \ref{sec:conc} summarizes our contributions and suggests directions for future work. 

All of our data and code to reproduce our results is hosted publicly at the first author's GitHub page: \url{https://github.com/nsandholtz/hotspot_paper}.

\section{Search Task Description} \label{sec:task_desc}

In collaboration with the Neuromotor Control Lab at Harvard University, we designed a ``hotspot" search task to present subjects with exploration vs exploitation conflicts while searching for an unknown optimal location.  On each round, subjects searched for an invisible target location (a ``hotspot") randomly placed on a 9-inch diameter circular task-region shown to them on a computer monitor.  Rounds consisted of 3 to 10 moves, where the number of moves was determined randomly according to a uniform distribution and the number was unbeknownst to the subject until the end of the round.  

Within a round, a ``move" entails sampling a location on the task region (i.e. clicking on it) after which a numerical score is immediately displayed to the user proportional to the click location's proximity to the hotspot.  Specifically, the reward received on move $t$ is a deterministic function of the distance from the click location to the hotspot: 
\begin{align}
 r_{t} &= f(d(\mathbf{m}_t)) \nonumber \\
         &= r_{0} + k \times (d(\mathbf{m}_{0}) - d(\mathbf{m}_{t})), \label{eq:objective_func}
\end{align}
where $r_{0}$ denotes the initial score, $d(\mathbf{m}_{t})$ denotes the euclidean distance (in pixels) of move $\mathbf{m}_t$ (a two-dimensional coordinate) to the hotspot, and $k$ denotes the reward scale (i.e. the reward's sensitivity to hotspot distance).  The reward scale $k$ is randomly generated for each new round of the task according to a Uniform[0, $5/3$] points/pixel distribution and is opaque to the subject.  This prevents subjects from gaining information about the objective function along the orthogonal direction of their first move, thus ensuring that the second move presents a tradeoff between exploration and exploitation.\footnote{If the reward scale was identical from round to round, subjects could solve for the gradient information along the direction orthogonal to their first move despite only having two responses from the objective, thus giving them a near-complete characterization of the local reward surface on their second move.  By randomizing the reward scale on each round, we ensure that subjects have uncertainty about the gradient information orthogonal to their first move. This argument implicitly assumes that we can linearize the response surface in the local region defined by the move 1 boundary.  We justify this assumption in Section \ref{sec:BO}.} 

\begin{figure}[H] 
\begin{center}
    \begin{subfigure}[b]{.32\textwidth}
       \includegraphics[trim={.8cm .2cm 0cm .8cm}, clip, width=1\textwidth]{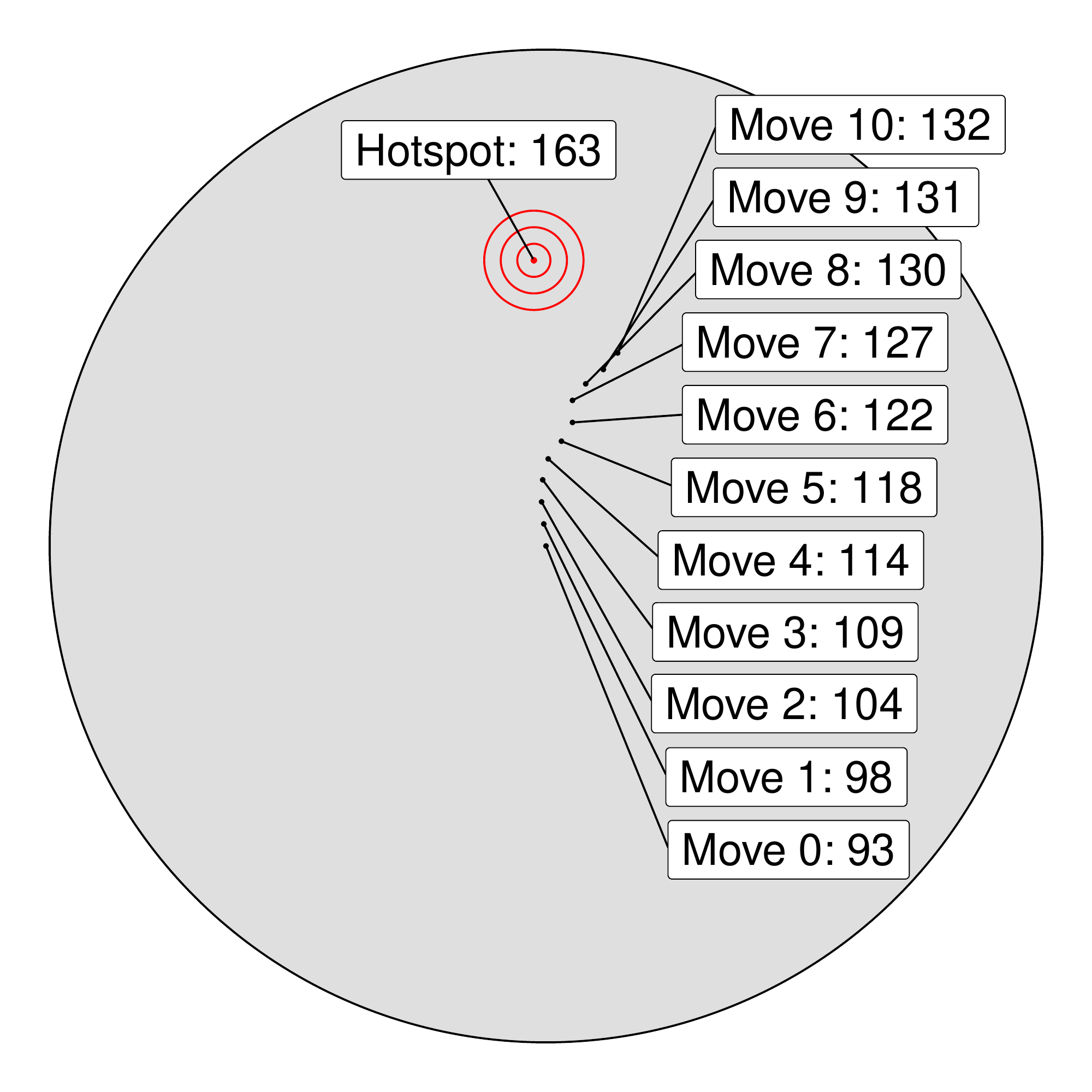}
        \caption{}
        \label{fig:full_round}
    \end{subfigure}  
    \begin{subfigure}[b]{.32\textwidth}
        \includegraphics[trim={4cm 0cm 4cm .6cm}, clip, width=1\textwidth]{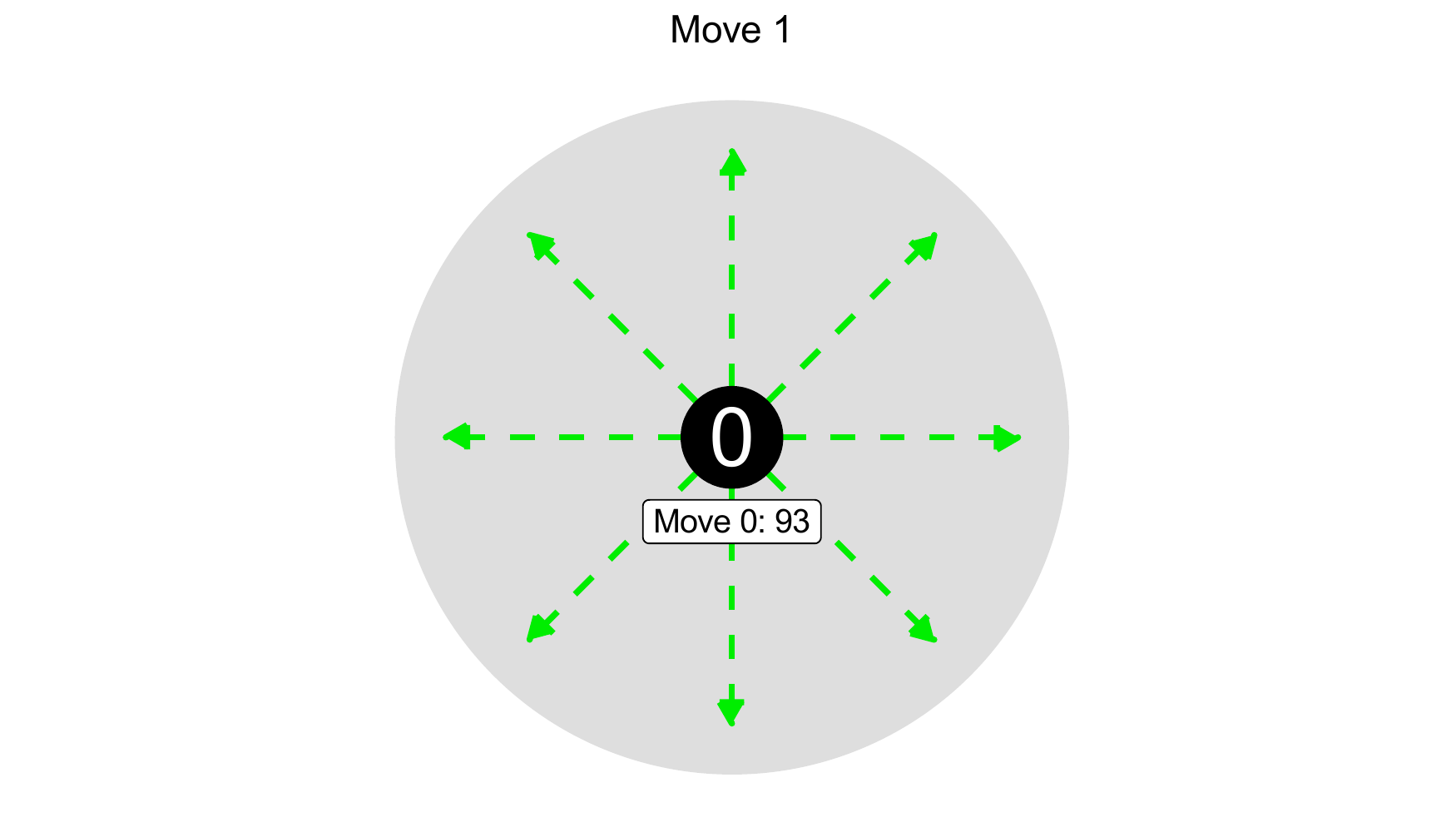}
        \caption{}
        \label{fig:move_1}
    \end{subfigure} 
    \begin{subfigure}[b]{.32\textwidth}
        \includegraphics[trim={5cm .7cm 5cm .6cm}, clip, width=1\textwidth]{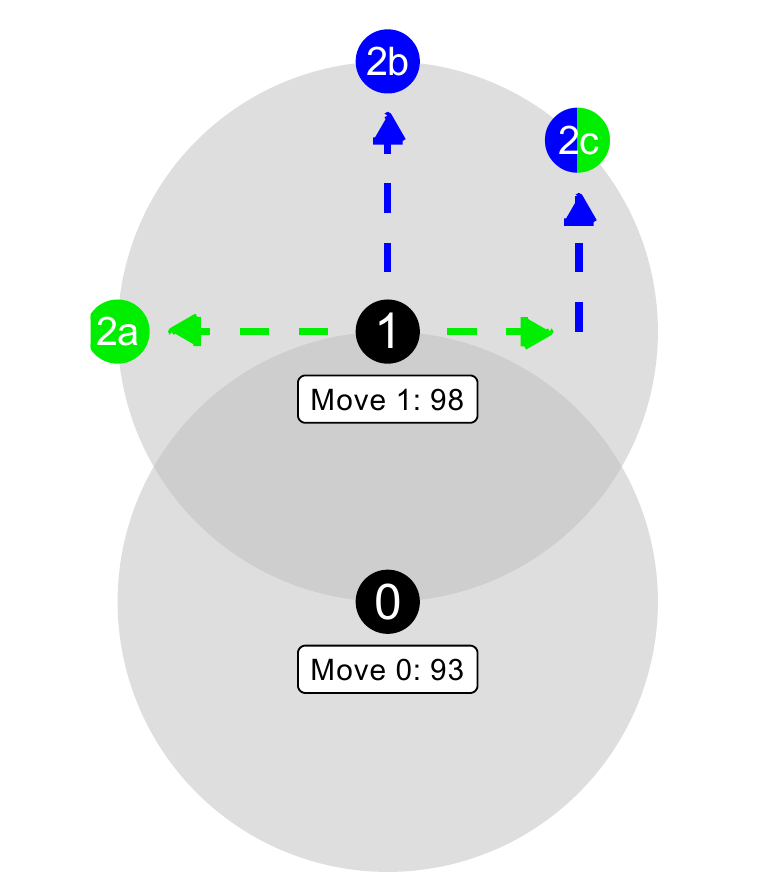}
        \caption{}
        \label{fig:move_2}
    \end{subfigure} 
\end{center}
\caption{(a): An example round of the hotspot task.  The red target shows the invisible hotspot location and the dots track the subject's search path.  Each move's score is shown in a text box.  The score at the hotspot is the score the subject would be given if they sampled the hotspot.  (b): Initial state (move 0) for the example round shown in (a), zoomed in to the move 1 click-region.  Moving in any direction represents exploration, as shown by the dashed green lines.  (c): The subject moves to the upper edge of the move 1 click-region, receiving a score of 98.  This creates a conflict between exploration (green) and exploitation (blue) on move 2.  Moving perpendicular to direction of move 1 represents pure exploration (move 2a), while moving along the same direction of move 1 represents pure exploitation (move 2b).  Any move between these extremes represents a combination of exploration and exploitation (move 2c).}
\label{fig:exp_conflict_diagram}
\end{figure}

To minimize effects of the task-region boundary guiding subject search behavior, the task always began at the center of the task region.  We refer to this initial starting point as ``move 0".  Each subsequent move was constrained to be within a small circular region (0.2 inch radius) around the previous move.  Subjects performed many rounds of this task; across the 28 subjects who participated, the minimum number of rounds played was 228 and the maximum was 716.  Panel (a) of Figure \ref{fig:exp_conflict_diagram} displays an example round of the experiment.  

We are primarily interested in the second move (move 2) of the task because this move provides the most information about a subject's exploration vs. exploitation preferences.  The first move (move 1) a subject makes is virtually random since they have no information about the direction of the hotspot upon beginning each round, as illustrated in panel (b) of Figure \ref{fig:exp_conflict_diagram}.  After receiving feedback from their first move, the subject acquires information about the gradient of the objective function \textit{but only along a single direction}. This poses a conflict between exploration and exploitation on the second move: continuing along the direction of their first move represents pure exploitation of their current knowledge, while any deviation from this direction represents some degree of exploration.  Thus the combination of a subject's first and second moves provides direct insight into how they balance exploration and exploitation, as illustrated in panel (c) of Figure \ref{fig:exp_conflict_diagram}. 

 In the next section we frame this task as a Bayesian optimization procedure.  Ultimately, our goal is to characterize and estimate subjects' exploration/exploitation preferences via the acquisition function of the Bayesian optimization model of their behavior on the first and second moves of the task. 

\section{Optimization Framework} \label{sec:BO}

Mathematically, the subjects' goal is to maximize an unknown
objective function $f$
\begin{equation}
\begin{aligned}
\underset{\mathbf{m} \in \mathcal{M}}{\max} \quad &  f(\mathbf{m})
\end{aligned} \label{eq:basic_problem}
\end{equation}
where $\mathbf{m}$ is a two-dimensional location and $\mathcal{M}$ is the circular task region defined on  $\mathbb{R}^2$.  However, as each move is constrained to be within a small region about the location of their previous move, the problem can be more precisely characterized as 
\begin{equation}
\begin{aligned}
\underset{\mathbf{m}_k \in \mathcal{M}_{k}}{\max} \quad & f(\mathbf{m}_k) \\
\textrm{s.t.} \quad & \mathbf{m}_0 = (0,0) \\
\quad & \mathbf{m}_i \in \mathcal{M}_{i}, ~~  i \in 1,\ldots,k 
\end{aligned} \label{eq:forward_problem}
\end{equation}
where $\mathcal{M}_{i}$ is the local click-region surrounding $\mathbf{m}_{i-1}$ as described in Section \ref{sec:task_desc}, and $k$ is randomly selected from the uniform distribution on the integers from 3 to 10.

In order to make inference on how the subject's solve \eqref{eq:forward_problem}, we model their optimization procedure as a Bayesian optimization \citep{borji2013bayesian}.  This entails characterizing the uncertainty about the unknown objective function with a surrogate, then strategically synthesizing the surrogate uncertainty via an acquisition function in order to determine promising  new  locations  to  sample.  We define this framework in the context of the first and second moves of the task, as we exclusively consider these moves in the inverse problem.  Appendix \ref{sec:move_3} in the supplemental material to this paper explains why moves 3 and up provide comparatively little information about the subjects' acquisition preferences.  

\subsection{Choosing a surrogate}

A `surrogate' is simply a term for a statistical model with an emphasis on pragmatism and prediction rather than interpretability and identification \citep{gramacy2020surrogates}.  Bayesian optimization applications typically use Gaussian process (GP) surrogates, which is a highly flexible class of models that can easily be updated as new samples from the objective are obtained.

\begin{figure}[H] 
\begin{center}
\includegraphics[trim={.5cm .1cm .5cm .1cm}, clip, width=.9\textwidth]{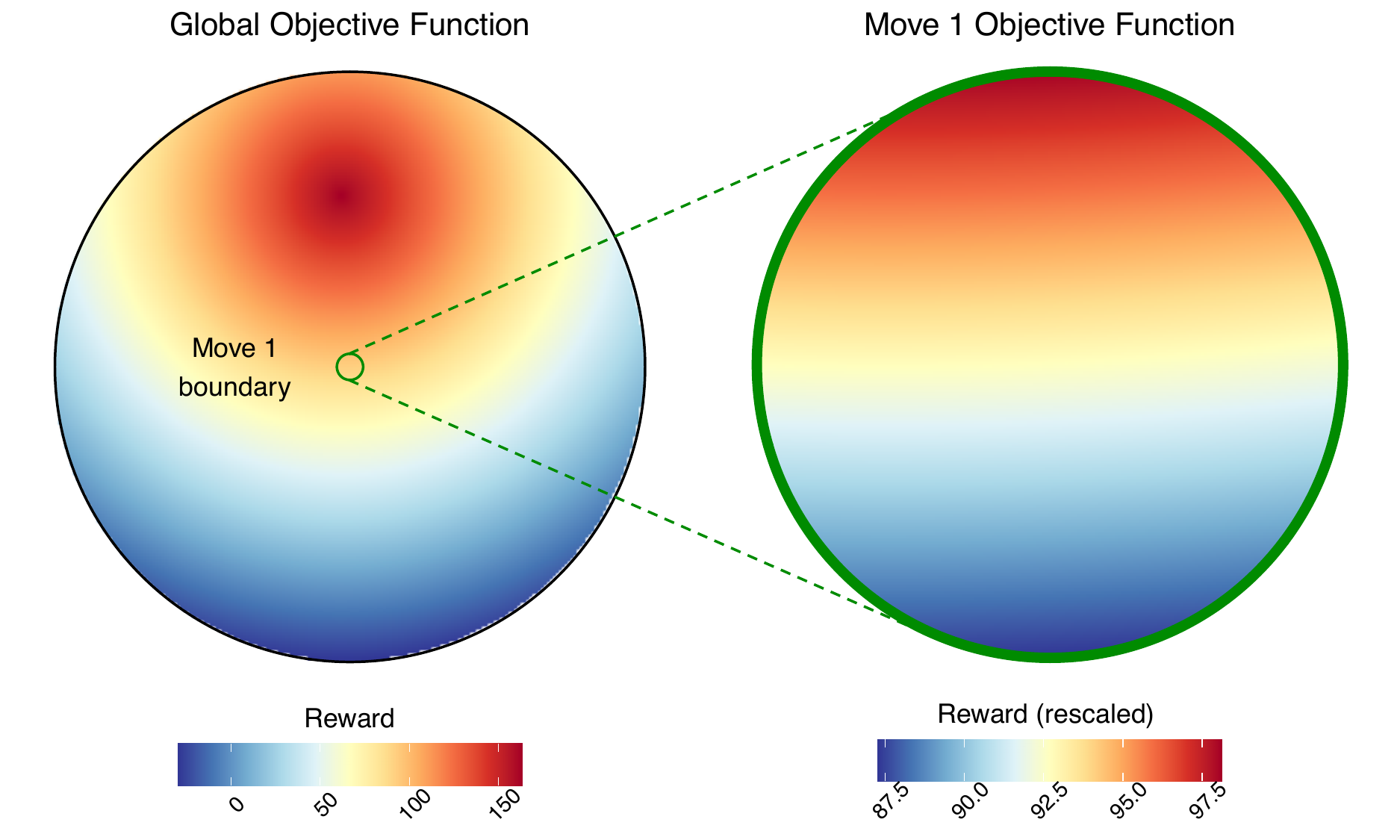}
\end{center}
\caption{\textit{Left}: The global objective function for the example round shown in Figure \ref{fig:exp_conflict_diagram}.  \textit{Right}: The objective function zoomed in to the click-region of the first move.}
\label{fig:reward_surface}
\end{figure}

In our case, GP surrogates would inaccurately specify the subjects' beliefs about the objective given their prior knowledge of the reward structure, and would therefore invalidate any inference about their acquisition function.  Subjects know a priori that there is an optimum somewhere on the circular task-region and that the rewards decrease uniformly and monotonically from this hotspot with distance. This reward structure is illustrated in Figure \ref{fig:reward_surface}, which shows the objective function from the example round shown previously. 

As illustrated in the left plot, globally the objective function is conical.  However, the experiments were designed to yield a surface that is approximately linear in the \textit{local} region around each individual move, as shown in the right panel.  Given the subjects' prior awareness of the reward structure, in addition to the approximate linearity in the neighborhood defined by a move's click-region, we assume that the subjects use a linear model as a surrogate of the objective: 
\begin{align}
\hat{f}_1(\mathbf{m}_{1}) & = r_{0} + \mathbf{m}_{1}^\intercal \boldsymbol{\beta} + \epsilon_{1}  \label{eq:surrogate}  \\
\epsilon_{1} &\sim \mathcal{N}(0,\sigma^2_{s}),  \nonumber
\end{align}
where $\mathbf{m}_{1}^\intercal = (x_{1}, y_{1})$, $\boldsymbol{\beta} = \icol{\beta^x \\ \beta^y}$ represents the reward gradient with respect to the Cartesian plane and $\epsilon_1$ represents a deviation from the surrogate to the true objective, which we assume to be Gaussian. We fix $\sigma_{s}$ at a tiny value ($\sigma_{s} = 0.01$) because the deviations from the linear model to the objective are negligible in the local region about $\mathbf{m}_{1}$.\footnote{In order to have an analytic update for the surrogate after each move, we use a Gaussian distribution to model the error term $\epsilon_1$.  While this greatly speeds up the inference when solving the inverse problem, it is actually a misspecification---the errors are not Gaussian.  Conditional on feedback from the first move, the model given by (\ref{eq:surrogate}) is effectively a tangent plane of the underlying conical objective, hence the $\epsilon_1$ are almost exclusively non-negative.  The mean of $\hat{\epsilon}_1$ across all rounds of the experiment is 0.006.}  

The approximate linearity of the objective function in the click-region around each move is an important feature of the experiment, as it is what allows us to assume that the directions of exploration and exploitation are orthogonal on move 2.  We note, however, that the fidelity of the linear approximation to $f$ declines as the distance to the hotspot decreases.  This fact required a careful design of the experiment; we wanted the majority of clicks to take place far from the hotspot to make the linear approximation more accurate, but we also wanted the hotspot to be reachable so that subjects performed the optimization seriously.  Balancing these competing desires, we selected the number of moves per round and the size of the local click-region such that it would be impossible to cover the distance to the hotspot in approximately 90\% of all rounds.  We omitted rounds where the hotspot was reachable in our analysis.   

\subsection{Updating the surrogate}

As subjects sample new locations and receive additional feedback, their uncertainty about the objective function diminishes.  In the context of Bayesian optimization, this corresponds to updating the surrogate as new information is gained about the objective function.  

\begin{figure}[h] 
\begin{center}
\includegraphics[trim={.5cm .8cm .25cm .15cm}, clip, width=.7\textwidth]{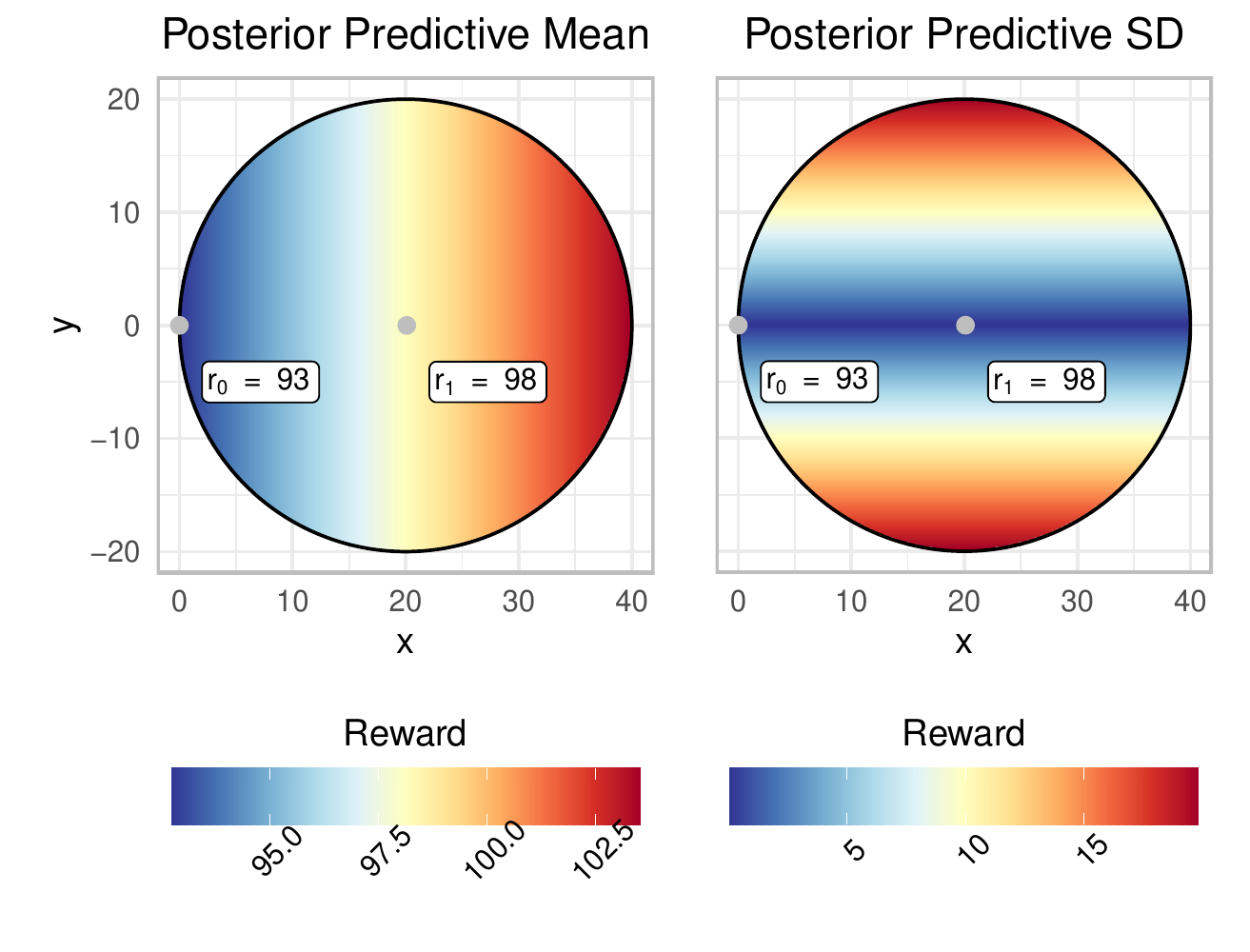}
\end{center}
\caption{The move 2 posterior predictive distribution of $\hat{f}_1$ given $r_0 = 93$ and $r_1 = 98$.  The left and right plots show the posterior predictive mean and standard deviation surfaces respectively, projected onto a fine grid over the move 2 click-region.}
\label{fig:example_posterior}
\end{figure}

We apply a Bayesian framework for the surrogate inference.  Since the hotspot is randomly placed over the task-region, we assume an isotropic, zero-mean Gaussian prior for $\boldsymbol \beta$:
\begin{align}
\bs \beta &\sim \mathcal{N}_2\big(\bs \mu_0 = \icol{0\\0}, \bs \Sigma_0 = \sigma^2_{\beta}\mathbf{I}\big), \label{eq:beta_prior} 
\end{align}
where $\sigma^2_{\beta}$ is a hyperparameter that we select to give a weakly-informative prior on the mean.  Let the observed search path after move $1$ be denoted as $\mathcal{D}_1 = \{\mathbf{M}_1, \mathbf{r}_1\}$, where $\mathbf{M}_1$ is a $2 \times 2$ matrix with the first row equal to (0,0) and the second row equal to $\mathbf{m}_1^\intercal$, and $\mathbf{r}_1$ is a lenth-2 column vector of observed rewards on moves 0 and 1.  Given $\mathcal{D}_1$,  (\ref{eq:beta_prior}) can be updated yielding a posterior distribution on $\bs \beta$.  The prior is conjugate and the posterior can be computed analytically  \citep[Online appendix]{taboga2021lectures}:
\begin{align}
\bs \beta | \mathcal{D}_1 &\sim \mathcal{N}_2(\bs \mu, \bs \Sigma) \label{eq:post_beta} \\
\bs \Sigma &=  \Big(\frac{1}{\sigma_{s}^2}\mathbf{M}_1^\intercal \mathbf{M}_1 + \bs \Sigma_0^{-1}\Big)^{-1}, \label{eq:post_sigma} \\
\bs \mu &= \bs \Sigma \Big(\frac{1}{\sigma_{s}^2}\mathbf{M}_1^\intercal(\mathbf{r}_1 - r_0) + \bs \Sigma_0^{-1} \bs \mu_0 \Big). \label{eq:post_mu}
\end{align}
The posterior predictive distribution of the surrogate for move 2 is given by:
\begin{align}
\hat{f}_{1}(\mathbf{m}_{2} | \mathcal{D}_1)  &\sim \mathcal{N}(\mathbf{m}_{2}^\intercal \bs \mu + r_0, ~ \mathbf{m}_{2}^\intercal \bs  \Sigma \mathbf{m}_{2} + \sigma_{s}^2), \label{eq:post_pred}
\end{align}
where $\mathbf{m}_{2}$ is a potential move 2 location and all other terms are as defined previously  \citep[Online appendix]{taboga2021lectures}. Figure \ref{fig:example_posterior} shows the updated surrogate predictive distribution for an example round after move 1.  

\subsection{Choosing an acquisition function}
The updated surrogate can be utilized to obtain promising new locations to sample.  This process is operationalized through the acquisition function.  The acquisition function $u$ is a function of a proposed location $\mathbf{m}_{2}$ and the surrogate after move 1 (which itself is a function of $\mathcal{D}_1$).  Acquisition functions are typically constructed such that high values of the function correspond to \textit{potentially} high values of the objective, either because the predicted mean is high, the uncertainty is high, or both \citep{brochu2010tutorial}.  The experimenter maximizes this function over $\mathcal{M}_{2}$, the set of potential move 2 locations, thus obtaining a new location to sample:  
\begin{align}
\mathbf{m}_{2} = \argmax_{\mathbf{m} \in \mathcal{M}_{2}}~ u(\mathbf{m} | \hat{f}_1, \mathcal{D}_1).
\label{eq:optimality_constraint}
\end{align}

A host of acquisition functions have been proposed in the literature (see \cite{shahriari2015taking} for a review).  While any acquisition function that selects new locations deterministically is a viable candidate for the methods we introduce in this paper, we restrict our analysis to probability of improvement (PI), expected improvement (EI), and an upper confidence bound (UCB) criterion since these functions are well-known among Bayesian optimization practitioners and have analytic solutions under Gaussian models \citep{jones1998efficient, srinivas2009gaussian}.  Mathematically, these are defined by 
 \begin{align}
 u_{\text{\tiny{PI}}}(\mathbf{m} | \hat{f}_1, \xi_{\text{\tiny{PI}}}, \mathcal{D}_1) &= \mathbb{P}\big(\hat{f}_1(\mathbf{m} | \mathcal{D}_1) \geq f(\mathbf{m}_1^+) + \xi_{\text{\tiny{PI}}}\big)  \label{eq:PI} \\
u_{\text{\tiny{EI}}}(\mathbf{m} | \hat{f}_1, \xi_{\text{\tiny{EI}}}, \mathcal{D}_1) &= \mathop{\mathbb{E}}[\text{max}⁡\big(0,\hat{f}_1(\mathbf{m} | \mathcal{D}_1) - (f(\mathbf{m}_1^+) + \xi_{\text{\tiny{EI}}})\big)] \label{eq:EI}\\
u_{\text{\tiny{UCB}}}(\mathbf{m} | \hat{f}_1, p, \mathcal{D}_1) &=  \text{infimum}\{\mathbf{m}: p \leq \hat{F}_1(\mathbf{m} | \mathcal{D}_1)\}, \label{eq:UCB}
 \end{align}
\noindent where $p > 0.5$, $\mathbf{m}_1^+$ is the best location observed so far over the two existing samples (i.e. $\mathbf{m}_1^+ = \text{argmax}_{\mathbf{m} \in \mathbf{M}_{1}} f(\mathbf{m})$), and $\hat{F}_1$ is the cumulative distribution function of $\hat{f}_1$.  

Each of the acquisition functions in (\ref{eq:PI})-(\ref{eq:UCB}) depend on an additional parameter which controls the premium on exploration.  Acquiring via PI \citep{kushner1964new} results in the location that most confidently predicts an increase in the response but ignores improvements less than $\xi_{\text{\tiny{PI}}}$.  EI \citep{mockus1978toward} considers the magnitude of improvement at a particular location, where $\xi_{\text{\tiny{EI}}}$ controls a tradeoff between exploration (high $\xi_{\text{\tiny{EI}}}$) and exploitation (low $\xi_{\text{\tiny{EI}}}$) \citep{lizotte2008practical}.  UCB acquisition functions \citep{cox1992statistical, srinivas2009gaussian} select new locations to sample based on optimistic estimates of the resulting reward at each location.  Higher values of $p$ encourage more exploration.\footnote{The UCB acquisition function is usually defined in terms of a mean function and covariance function since the surrogate is a Gaussian process in most Bayesian optimization applications.  We define it via the quantile function in order to have an upper and lower bound on the parameter $p$, which simplifies the methodology in Section \ref{sec:IBO}.}   

\begin{figure}[H] 
\begin{center}
\includegraphics[trim={0cm 0cm 0cm 0cm}, clip, width=1\textwidth]{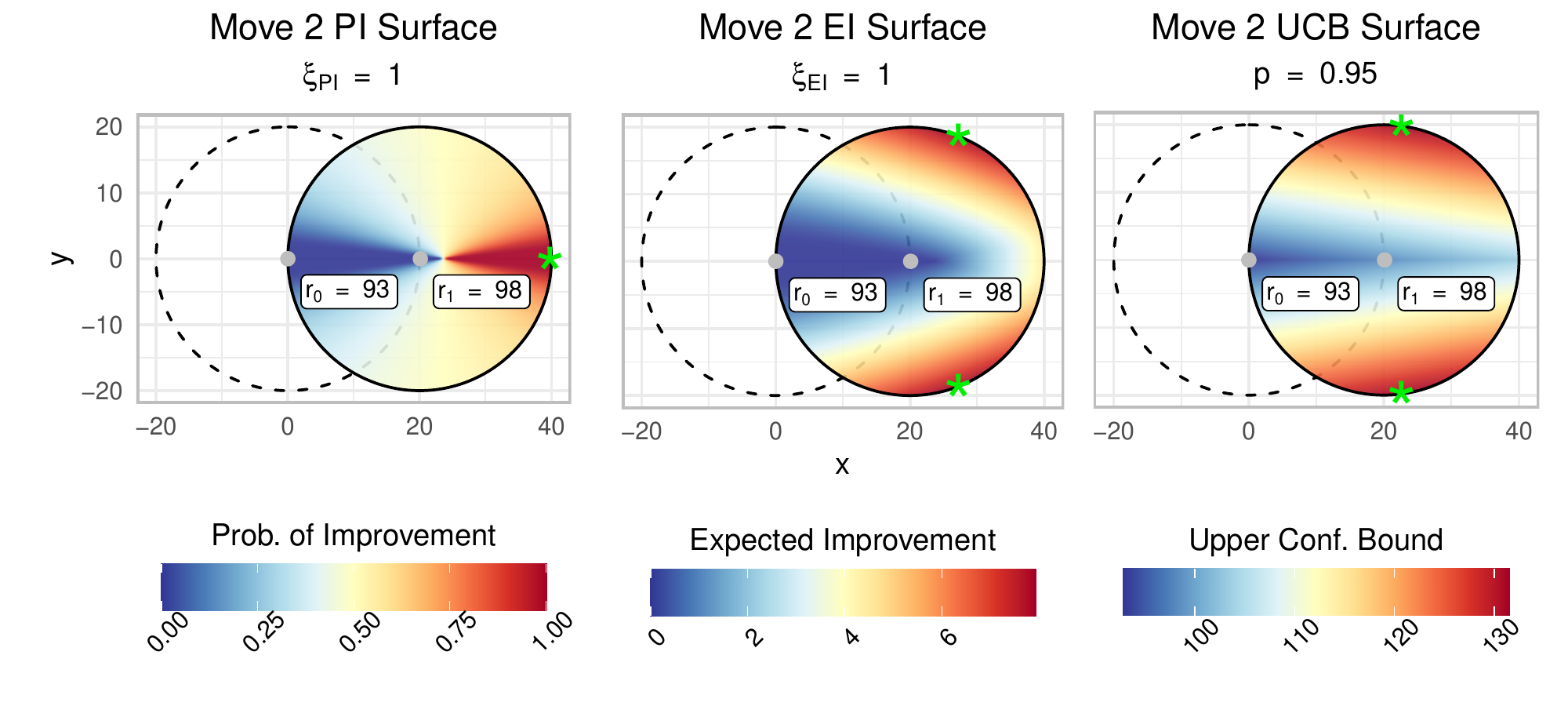} 
\end{center}
\caption{Acquisition surfaces for the PI, EI, and UCB functions defined in (\ref{eq:PI})-(\ref{eq:UCB}) given $r_0 = 93$ and $r_1 = 98$.  Each surface is on a different scale: the PI surface (left) is on the probability scale, EI (middle) is in terms of points over 99, and UCB (right) is in terms of the 95th percentile of the surrogate.  The click-region boundary is shown by the black circle encompassing the colored surfaces and the $\argmax$(s) of each surface is denoted by a green star.}
\label{fig:example_acquisition_m2}
\end{figure}

Figure \ref{fig:example_acquisition_m2} shows the acquisition surfaces for the PI, EI, and UCB functions defined in (\ref{eq:PI})-(\ref{eq:UCB}) in the context of the example round shown in Figure \ref{fig:exp_conflict_diagram}.  We reoriented the data so that move 1 falls along the horizontal-axis for visual clarity.  The $\argmax$(s) of each surface is denoted by a green star.

Notice that each method prescribes different move 2 locations; PI recommends pure exploitation, while EI and UCB recommend almost pure exploration.  Of course, this is not always the case.  Depending on the change in score from move 0 to move 1, in addition to the exploration parameter values ($\xi_{\text{\tiny{PI}}}$, $\xi_{\text{\tiny{EI}}}$, and $p$), the optimal locations returned by the acquisition functions can vary substantially.  Also note that each surface is symmetric across the exploitation axis (i.e. $y=0$).  This is a feature of the subject having gained information exclusively about a single direction after the first move.  Due to this phenomenon, the surfaces may be bimodal or unimodal.  

Because we use a linear model as the surrogate for the objective function, the optimal location to sample on move 2 (as determined by \eqref{eq:optimality_constraint}) always falls on the click-region boundary, regardless of the change in score on move 1.  This harmonizes with the subjects' actual behavior; in 95\% of the rounds, subjects moved to the click-region  boundary on move 2.\footnote{The search task was programmed to snap the cursor to the edge of the boundary if the mouse exceeded it, which made it easy for subjects to make moves along the boundary.}  This effectively allows us to reduce the $\argmax$ search from two dimensions to one dimension.  Leveraging polar coordinates, we can fix the radius of the move 2 location at the click-region boundary and limit the $\argmax$ search solely to $\theta_2$, the angle of move 2 relative to the direction of move 1:
\begin{align}
\argmax_{\mathbf{m} \in \mathcal{M}_2}~ u(\mathbf{m} | \hat{f}_1, \mathcal{D}_1) \longrightarrow \argmax \limits_{\theta \in (-\pi, \pi]} u(\theta | \hat{f}_1, \mathcal{D}_1). \nonumber
\end{align}  

\begin{figure}[H] 
\begin{center}
\includegraphics[trim={0cm .6cm 0cm .15cm}, clip, width=1\textwidth]{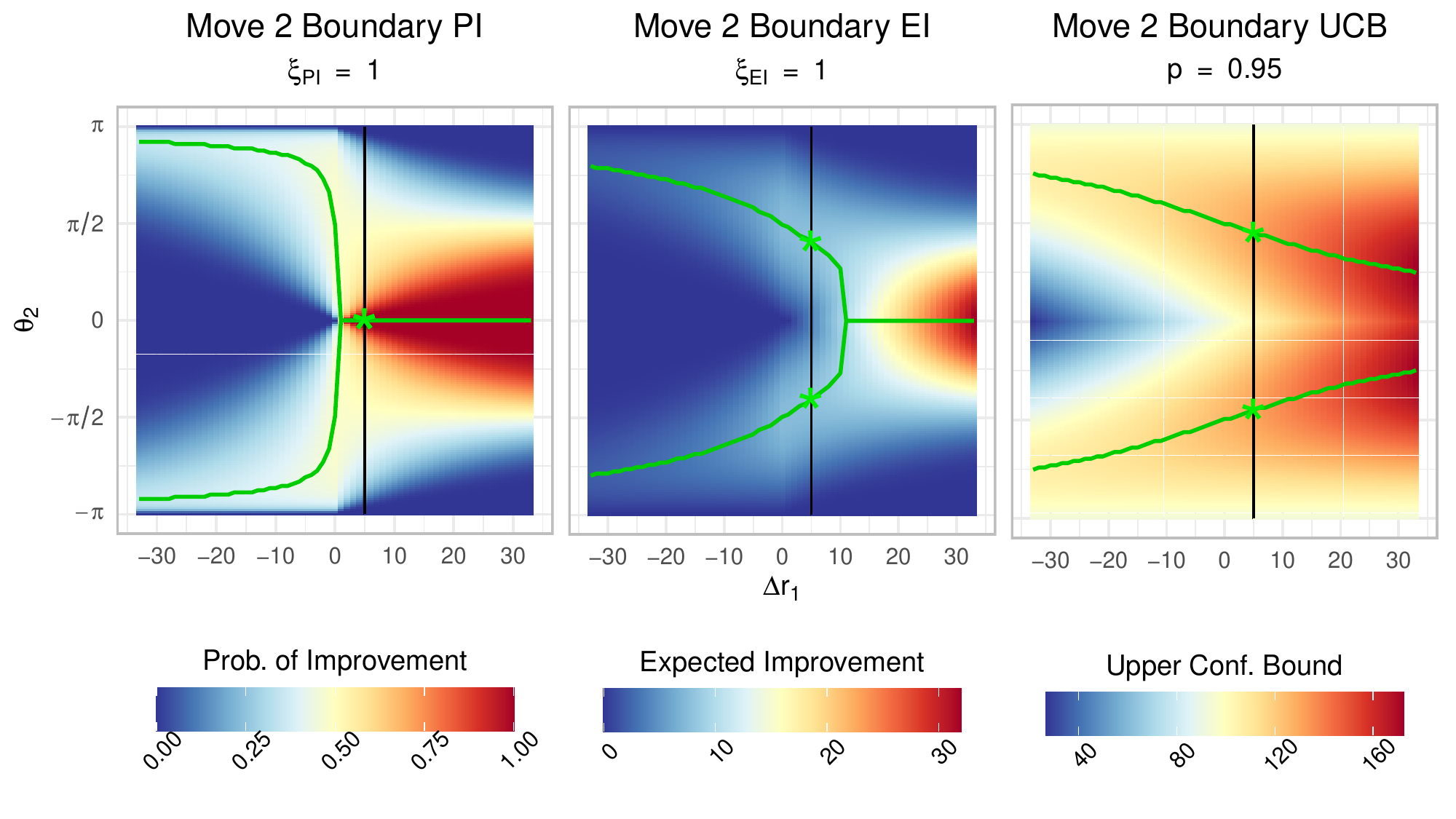}
\end{center}
\caption{Click-region boundary acquisition values over the range of possible $\Delta r_1$ values for three sample acquisition functions.  The horizontal axis denotes $\Delta r_1$, the change in reward from $\mathbf{m}_0$ to $\mathbf{m}_1$.  The vertical axis shows $\theta_2$, the angle of the second move relative to the first move.  Color indicates the acquisition function value for any given ($\Delta r_1, \theta_2$) pair.  In each plot, the green curve denotes the angle that yields the maximum of the boundary acquisition values as a function of $\Delta r_1$.  The vertical black lines at $\Delta r_1 = 5$ correspond to the circular black lines denoting the click-region boundaries in Figure \ref{fig:example_acquisition_m2}.  Similarly, the green stars correspond to the stars in Figure \ref{fig:example_acquisition_m2}.}
\label{fig:example_acquisition_curves}
\end{figure}

By reducing the dimensionality of the optimization we can illustrate acquisition values over the entire range of $\Delta r_1 = r_1 - r_0$, the change in score after a subject's first move, as shown in Figure \ref{fig:example_acquisition_curves}.  The three plots show acquisition values over a fine grid of ($\Delta r_1, \theta_2$) pairs for the same acquisition functions shown in Figure \ref{fig:example_acquisition_m2}.   In each plot, the green curves denote the angles that yield the maximum of the acquisition values as a function of $\Delta r_1$. Note that the move 2 locations are assumed to be made on the click-region boundary. 

\begin{figure}[H]
    \begin{center}
        \includegraphics[trim={0cm .25cm 0cm .15cm}, clip, width=.95\textwidth]{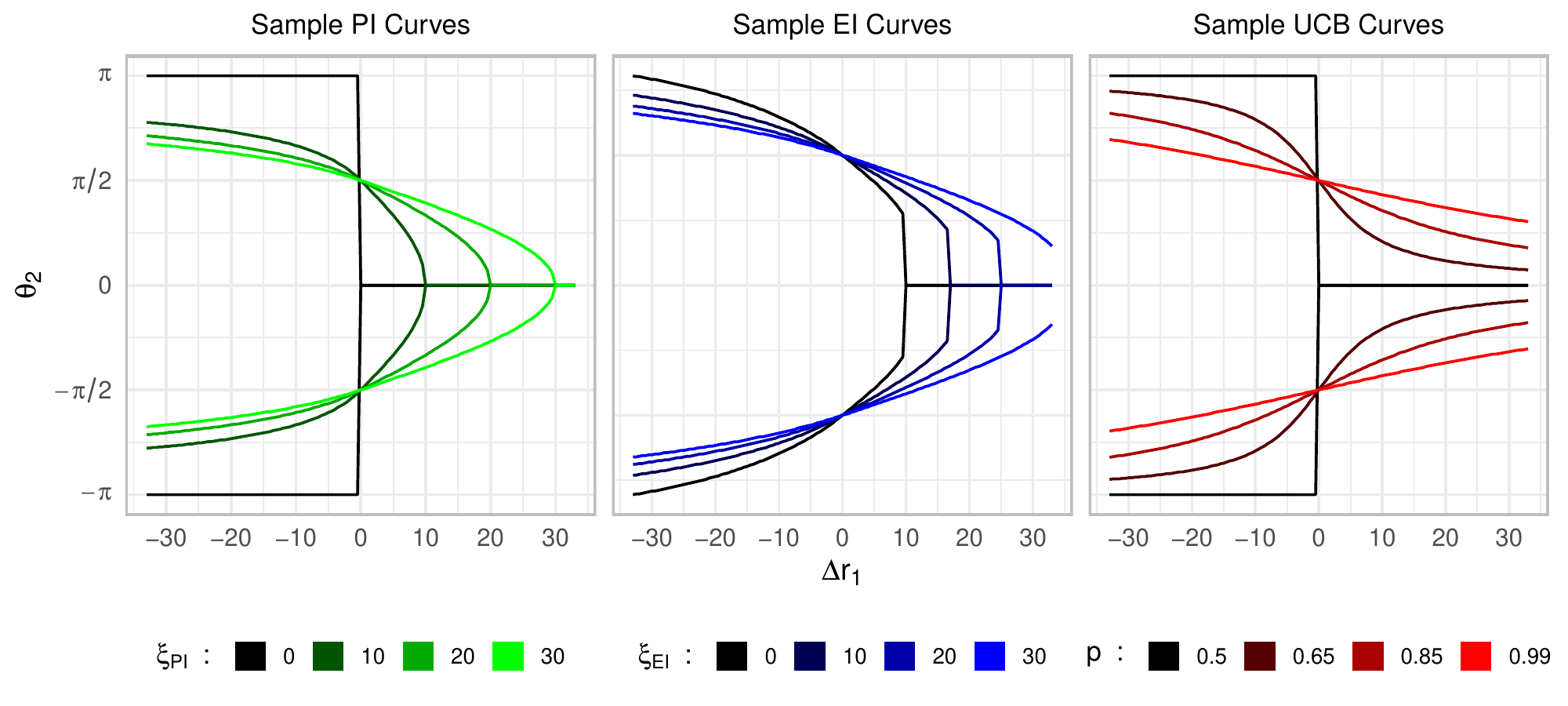}
      \end{center}
    \caption{Acquisition curves for various exploration parameter values of the corresponding family---PI (left), EI (middle), and UCB (right).} \label{fig:sample_acquisition_curves}
\end{figure}

Going forward, we will refer to an \textit{acquisition curve} as the curve defined by the optimal $\theta_2$ as a function of $\Delta r_1$ for a given acquisition function.  Figure \ref{fig:sample_acquisition_curves} shows various acquisition curves from the PI, EI, and UCB acquisition families.  Depending on the acquisition family and exploration parameter, the acquisition curves can vary substantially.  

Our goal is not to determine an optimal strategy, or acquisition function, for this task.  Rather, using Bayesian optimization as a model of each subject's optimization procedure, we aim to learn subject-specific acquisition functions guided by their observed behavior in the experiment.  

Before proceeding to Section \ref{sec:IBO}, which formally introduces the inverse problem, we pause to comment on the unconventional nature of the Bayesian optimization framework we have developed thus far.  The reward structure and transparent experimental setup (which allows subjects to understand the reward surface in relatively few moves) were carefully designed to allow us to assume that subjects' beliefs about the reward surface can be approximated by a linear surrogate.  This is crucial in enabling us to decouple uncertainty about a subject's surrogate from uncertainty about their acquisition function, thus allowing us to make plausible inference on their acquisition function. 

\section{Learning Human Acquisition Functions} \label{sec:IBO}

\begin{figure}[H]
    \begin{center}
        \includegraphics[trim={0cm .25cm 0cm .15cm}, clip, width= 1\textwidth]{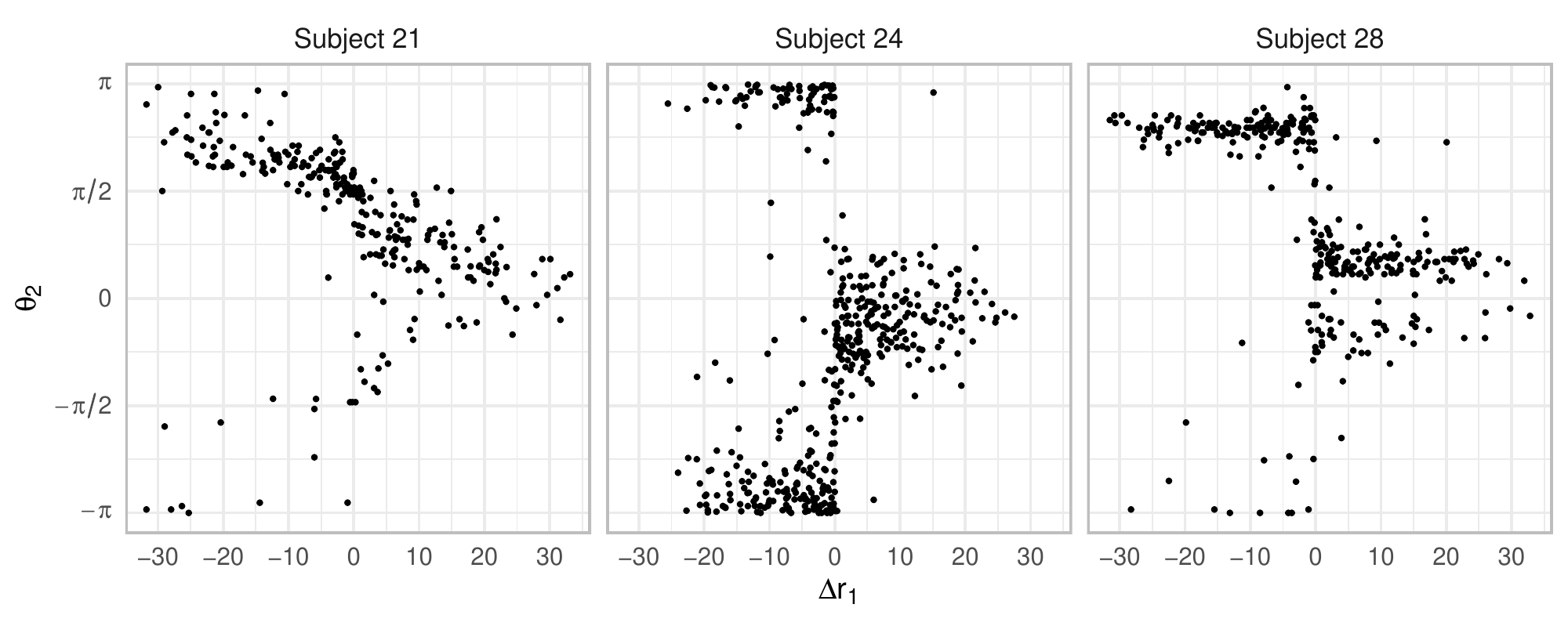} 
      \end{center}
    \caption{Move 2 behavior for subjects 21 (left), 24 (middle), and 28 (right).  Each scatterplot shows the ($\Delta r_1$, $\theta_2$) pairs for the subject's rounds of the task.} \label{fig:subject_ee_data}
\end{figure}

In this section we propose a method to estimate a subject's unknown acquisition function $u$ given their observed decisions on move 2 assuming that these decisions are approximately optimal solutions to \eqref{eq:optimality_constraint} with respect to $u$.  We say `approximately optimal' because the subjects' decisions exhibit considerable variability, as illustrated in Figure \ref{fig:subject_ee_data} which shows the move 2 behavior for three subjects in our study as a function of $\Delta r_1$.\footnote{Figure \ref{fig:move_2_xy} in Appendix \ref{sec:additional_figures} of the supplemental material shows the ($x$,$y$) coordinates of these same data.}
 This variability could arise from a number of factors.  For one, the modeling described in Section \ref{sec:BO} happens subconsciously;  subjects ``update" their uncertainty about the objective function intuitively, which introduces human error.  Noise could also be due to variation in a subject's ability to click exactly where they intend to, changes to their sampling strategy, or sloppiness from performing the task rapidly.  

 Formally, given a collection of $N$ move 2 angles $\boldsymbol \theta_\mathbf{2} = (\theta_2^1, \ldots, \theta_2^N)$ that were taken after receiving feedback from the corresponding move 1 (as represented by $\boldsymbol \Delta \mathbf{r_1} = (\Delta r_1^1, \ldots, \Delta r_1^N)$, the inverse problem we consider is to identify an acquisition function $u \in \mathcal{U}$ that minimizes a given loss function $\ell(\cdot,\cdot)$ 
\begin{equation}
    \min_{u \in \mathcal{U}} \ell(\boldsymbol \theta_\mathbf{2}, \boldsymbol \theta_\mathbf{2}^*(\boldsymbol \Delta \mathbf{r_1}, u)), 
\end{equation} \label{eq:approx_inv}
where $\boldsymbol \theta_\mathbf{2}^*(\boldsymbol \Delta \mathbf{r_1}, u) = (\theta_2^*(\Delta r_1^1, u), \ldots, \theta_2^*(\Delta r_1^N, u))$ is the vector of optimal decisions with respect to $u$ conditional on  $\boldsymbol \Delta \mathbf{r_1}$, with components defined as
\begin{equation}
\theta_2^*(\Delta r_1^i, u) = \argmax \limits_{\theta \in (-\pi, \pi]} u(\theta | \Delta r_1^i, \hat{f}_1) \label{eq:theta_2}
\end{equation}
for $i = 1, \ldots, N$. The function class $\mathcal{U}$ contains candidate acquisition functions that summarize the value of a particular move 2 angle for a given $\Delta r_1$ value and surrogate $\hat{f}_1$.  

Conceptually, solving (\ref{eq:approx_inv}.1) can be understood as finding acquisition curves (see Figure \ref{fig:sample_acquisition_curves}) that best fit the humans' observed optimization behavior (see Figure \ref{fig:subject_ee_data}).   As such, the problem becomes one of inference rather than optimization.  The optimization has already occurred---we want to infer \textit{how} each subject optimized. 

\subsection{A probabilistic solution framework}\label{sec:imperfect_acquisition}

We assume subjects' optimization behavior is characterized by an average strategy but that their behavior deviates randomly around this strategy from round to round.   Under this assumption, we propose a probabilistic solution framework for (\ref{eq:approx_inv}.1) as follows:

\begin{enumerate}
\item Assume a likelihood for ($\theta_{2}^i | \Delta r_1^i, \hat{f}_1, u$), parameterized such that the mode of the distribution equals the $\argmax$ of the acquisition function $u$.    
\item Select a set of candidate acquisition functions $\mathcal{U}$ as potential characterizations of the optimizer's preferred strategy and set a prior distribution over $\mathcal{U}$. 
\item Compute the posterior probability of each candidate acquisition function $u \in \mathcal{U}$:
\begin{align}
 p(u | \boldsymbol \theta_{\mathbf{2}}, \boldsymbol \Delta \mathbf{r_1}, \hat{f}_1) \propto  \prod_{i = 1}^N \ell \left(\theta_2^i \,\middle\vert\, \argmax_{\theta \in (-\pi, \pi]} u(\theta | \Delta r_1^i, \hat{f}_1)\right) p(u). \label{eq:posterior_ibo}
\end{align}   
\end{enumerate}
where $\ell(\cdot|\cdot)$ is the likelihood from step 1, and $p(\cdot)$ is the prior from step 2.  

We employ a Bayesian framework in our approach, but the procedure could be carried out using other estimators (e.g. maximum likelihood).  We now explain how we implement each of these steps in context of the hotspot search task.

\subsubsection{Choosing the likelihood}
  To model the symmetric bimodal nature of our data, we assume $\theta_2^i$ follows a two-component wrapped Cauchy mixture distribution:\footnote{When only a single value defines $\argmax_{\theta \in (-\pi, \pi]} u(\theta | \Delta r_1^i, \hat{f}_1)$, the likelihood simplifies to a single wrapped Cauchy distribution.}  
\begin{align}
\ell(\theta_2^i |  w, \Delta r_1^i, u, \hat{f}_1, \gamma) &= w \times h^-(\theta_2^i |  \Delta r_1^i, u,  \hat{f}_1, \gamma) + (1 - w) \times h^+(\theta_2^i |  \Delta r_1^i, u,  \hat{f}_1, \gamma), \label{eq:symmetric_distribution}
\end{align}
where
\begin{align}
h^-(\theta_2^i |  \Delta r_1^i, u,  \hat{f}_1, \gamma) &= \sum \limits_{k = -\infty}^{\infty} \frac{\gamma}{\pi(\gamma^2 + (\theta_2^i - \argmax \limits_{\theta \in (-\pi, 0]} u(\theta | \Delta r_1^i, \hat{f}_1) + 2\pi k)^2)}, \label{eq:wrapped_cauchy_pos} \\
h^+(\theta_2^i |  \Delta r_1^i, u,  \hat{f}_1, \gamma) &= \sum \limits_{k = -\infty}^{\infty} \frac{\gamma}{\pi(\gamma^2 + (\theta_2^i - \argmax \limits_{\theta \in (0, \pi]} u(\theta | \Delta r_1^i, \hat{f}_1) + 2\pi k)^2)}. \label{eq:wrapped_cauchy_neg}
\end{align}
The location of each mixture component in \eqref{eq:symmetric_distribution} is constrained to be at the argmax(s) defined by $\argmax_{\theta \in (-\pi, \pi]} u(\theta | \Delta r_1^i, \hat{f}_1)$.  The parameter $\gamma$ measures a subject's variation about their acquisition curve and the mixture weight $w$ governs a subject's propensity to favor exploring to the right over exploring to the left.   We use the wrapped Cauchy distribution because it is more robust to outliers, which are fairly common across many subject's move 2 behavior. 

Together, (\ref{eq:symmetric_distribution})-(\ref{eq:wrapped_cauchy_neg}) yield a likelihood that adheres to the defining features of our data.  It is not only defined on the proper support for $\theta_2$ (i.e. $(-\pi, \pi]$), it is also guaranteed to be symmetric over the axis of exploitation (i.e. $\theta_2 = 0$); since the argmaxes defined in the denominators of \eqref{eq:wrapped_cauchy_pos} and \eqref{eq:wrapped_cauchy_neg} will always be reflected about $\theta = 0$, the corresponding mixture distribution will also be symmetric about $\theta = 0$.  

The fact that we parameterize the likelihood such that the mode equals the $\argmax$ of the acquisition function is what renders this an `inverse optimization' method.  The novelty is that we couch this restriction within a likelihood, thereby allowing subject behavior to deviate around an `optimal' strategy (with respect to a given acquisition function $u$) according to the specified probability distribution.   This combination distinguishes our approach from both from other probabilistic models and other inverse optimization methods. 

\subsubsection{Setting priors}
We select the PI, EI, and UCB acquisition families defined in (\ref{eq:PI})-(\ref{eq:UCB}) as the set of candidate acquisition functions $\mathcal{U}$.  Note that our method does not preclude using other acquisition families---any acquisition function can be included in the candidate set $\mathcal{U}$.  Each acquisition function $u \in \mathcal{U}$ is assumed to have equal prior probability, thus we place a uniform prior over the parameters indexing the acquisition function space: $\xi_{\text{\tiny{PI}}}$, $\xi_{\text{\tiny{EI}}}$, $p$.   We also put a uniform prior distribution over $\gamma$, the scale parameter governing the variability in each subject's optimization behavior.  Appendix \ref{sec:prior_specification} in the supplemental material to this paper contains additional details on prior specification. 

\subsubsection{Approximating the posterior}
The existence of the $\argmax$ in the parameterization of (\ref{eq:symmetric_distribution})-(\ref{eq:wrapped_cauchy_neg}) creates an additional optimization step when evaluating (\ref{eq:posterior_ibo}).  To ease the computational burden posed by this additional optimization subroutine, we precompute a fine grid of acquisition curves over a plausible range of values for each candidate family and restrict our inference to this discrete set.  We also precompute the maximum likelihood estimate of the mixture weight $w$ using the EM algorithm \citep{dempster1977maximum}.  We can then compute  (\ref{eq:posterior_ibo}) for each $u \in \mathcal{U}$ by multiplying its prior probability by the joint likelihood of $\boldsymbol \theta_{\mathbf{2}}$ (plugging in the precomputed $\hat{w}$) and the prior over $\gamma$.  

\begin{figure}[H]
\begin{center}
\includegraphics[trim={.5cm .4cm 0cm 0cm}, clip, width=1\textwidth]{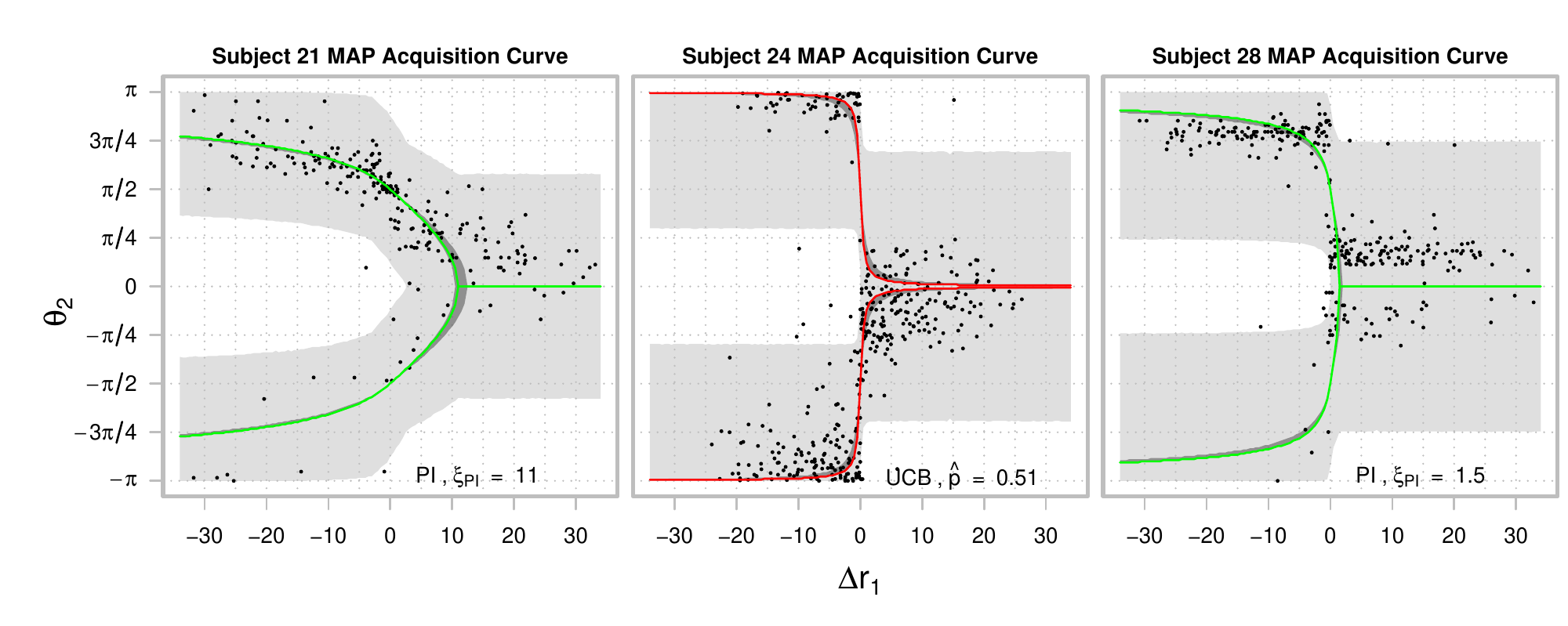} 
\end{center}
\caption{($\Delta r_1$, $\theta_2$) pairs for subjects 21 (left), 24 (middle), and 28 (right) overlaid with each subject's corresponding MAP acquisition curve.  In each plot, text in the lower-right corner shows the estimated exploration parameter for the subject's MAP acquisition function.  Around each curve is an approximate 95\% HPD credible region on the acquisition curve (dark-gray) and 95\% HPD prediction region (light-gray).}
\label{fig:normal_acquisition_fits}
\end{figure}

Figure \ref{fig:normal_acquisition_fits} shows the ($\Delta r_1$, $\theta_2$) pairs for subjects 21, 24, and 28 overlaid with their corresponding maximum a posteriori (MAP) acquisition curves in color.  The light-gray shaded areas show approximate 95\% highest posterior density (HPD) prediction regions, while the dark-gray regions denote 95\% HPD credible regions on the acquisition curves.  
 
\subsection{Incorporating human tendencies}
On visual inspection of Figure \ref{fig:normal_acquisition_fits}, the MAP acquisition curves for subjects 21 and 24 fit their data fairly well but best fit curve for subject 28 fits poorly.  This is because there are no candidates in $\mathcal{U}$ that prescribe such a drastic change in strategy from negative $\Delta r_1$ to positive $\Delta r_1$ values while maintaining highly exploratory preferences.  The set of acquisition functions as currently defined cannot yield acquisition curves with this shape.  Consequently, there is no parameter setting of $\xi_{\text{\tiny{PI}}}$, $\xi_{\text{\tiny{EI}}}$, or $p$ which renders subject 28's average behavior as optimal.

In order to allow this type of behavior to be rendered optimal, we propose augmenting the acquisition functions by an additional parameter $\tau$.  For a given acquisition function $u$, we define augmented acquisition function $\widetilde{u}$ as
\begin{align}
 \widetilde{u} (\theta_2 | \Delta r_1,\hat{f}_1, \tau) &= \begin{cases} u(\theta_2 | \Delta r_1, \hat{f}_1) &\mbox{if }   (|\theta_2| > \tau) \cap (\Delta r_1 \geq 0 ), \\
 u(\theta_2 | \Delta r_1, \hat{f}_1) &\mbox{if }   (|\theta_2| <  \pi - \tau) \cap ( \Delta r_1 < 0), \\
\min \limits_{\theta_2, \Delta r_1}(u(\theta_2 | \Delta r_1, \hat{f}_1)) &\mbox{otherwise}, 
\end{cases}  \label{eq:threshold}
\end{align}
where $\tau \in [0, \pi/2]$. Equation (\ref{eq:threshold}) defines the augmented value to be the minimum acquisition value (w.r.t. $\theta_2$ and $\Delta r_1$) for all move 2 angles that aren't at least $\tau$-radians exploratory, either in the forward or backward directions.  If a given move 2 angle is greater than $\tau$ and less then $\pi - \tau$ (in absolute value), the move is ``sufficiently exploratory" and the acquisition value is unchanged.  

\begin{figure} 
\begin{center}
\includegraphics[trim={0cm .25cm 0cm .15cm}, clip, width=1\textwidth]{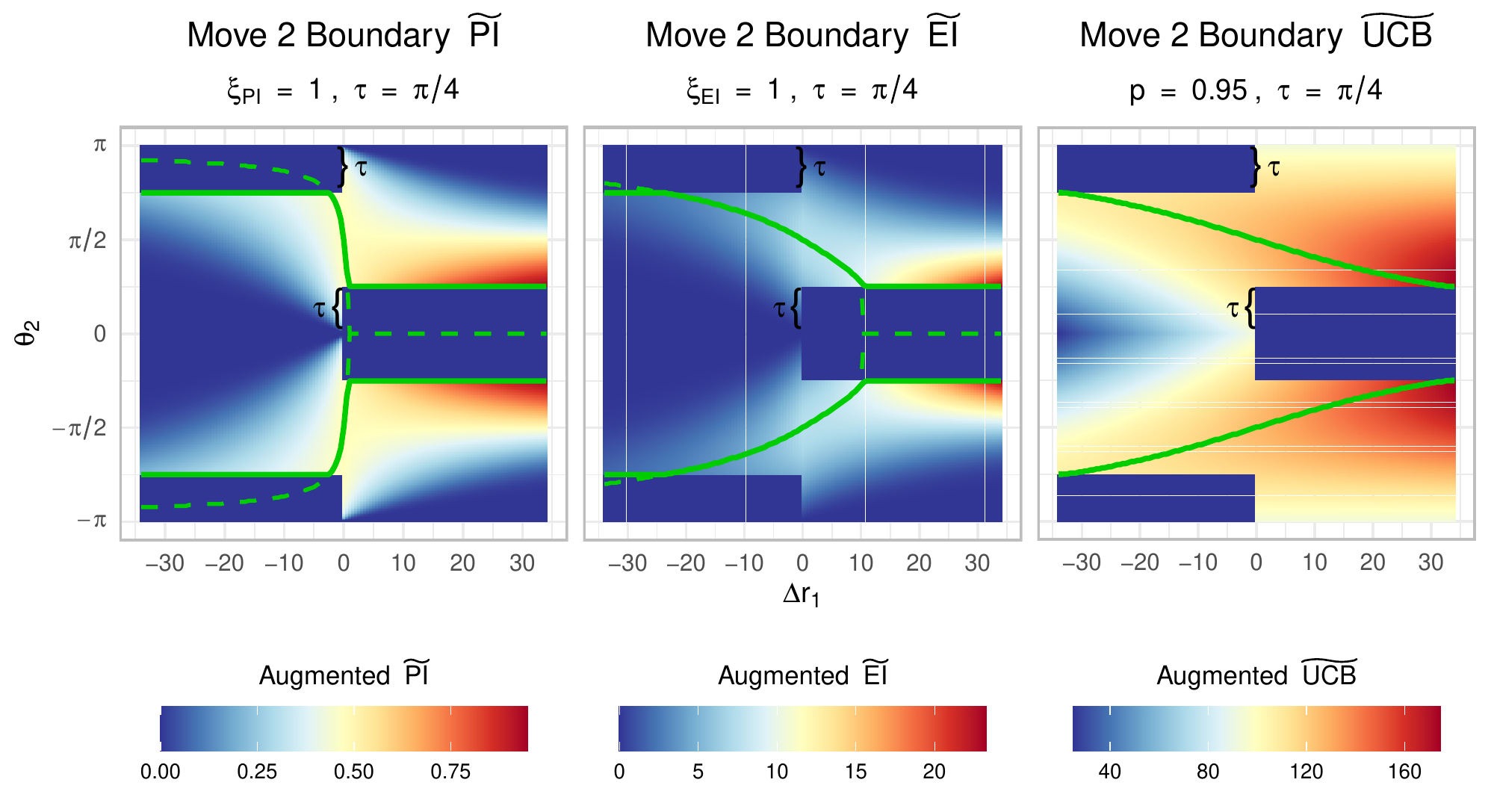} 
\end{center}
\caption{Click-region boundary augmented acquisition values over the range of possible $\Delta r_1$ values for three sample acquisition functions.  The solid green curves denote the angles that yield the maximum of $\widetilde{u} (\theta_2 | \Delta r_1,\hat{f}_1, \tau)$, while the dashed green curves show the angles that yield the maximum of $u$ without being augmented by $\tau$ (i.e. $\argmax u(\theta_2 | \Delta r_1, \hat{f}_1)$).}
\label{fig:tau_comparison}
\end{figure}

Figure \ref{fig:tau_comparison} illustrates how this augmentation changes the acquisition surfaces and corresponding acquisition curves for the three acquisition functions represented previously in Figure \ref{fig:example_acquisition_curves}.  

Another feature we account for is the human tendency to react differently for positive vs. negative feedback \citep{tversky1979prospect}.  Some subjects exhibit different exploration tendencies depending on whether they get a negative or positive change in score on their first move.  In order to allow for this type of behavior we modify (\ref{eq:threshold}) to allow different values of $\tau$ depending on whether $\Delta r_1$ is positive or negative: 
\begin{align}
 \widetilde{u}^\pm (\theta_2 | \Delta r_1,\hat{f}_1, \tau^+, \tau^-) &= \begin{cases} u(\theta_2 | \Delta r_1, \hat{f}_1) &\mbox{if }   (|\theta_2| > \tau^+) \cap (\Delta r_1 \geq 0 ), \\
 u(\theta_2 | \Delta r_1, \hat{f}_1) &\mbox{if }   (|\theta_2| <  \pi - \tau^-) \cap ( \Delta r_1 < 0), \\
\min \limits_{\theta_2, \Delta r_1}(u(\theta_2 | \Delta r_1, \hat{f}_1)) &\mbox{otherwise}.
\end{cases} \label{eq:threshold_pm}
\end{align}
 
The practical effect of \eqref{eq:threshold}-\eqref{eq:threshold_pm} is that they allow the optimization criteria to be based solely on exploration.  While the parameters in the unmodified PI, EI, and UCB acquisition families allow the optimizer to \textit{balance} exploration vs. exploitation differently when synthesizing the uncertainty in $\hat{f}$, they do not enable the optimizer to let exploration \textit{completely dominate} exploitation.  The augmented acquisition functions allow exploration to trump exploitation, regardless of $\hat{f}_1$.  We observe this type of behavior by many subjects in our study.  

\begin{figure}[h] 
\begin{center}
\includegraphics[trim={.5cm .4cm 0cm 0cm}, clip, width=1\textwidth]{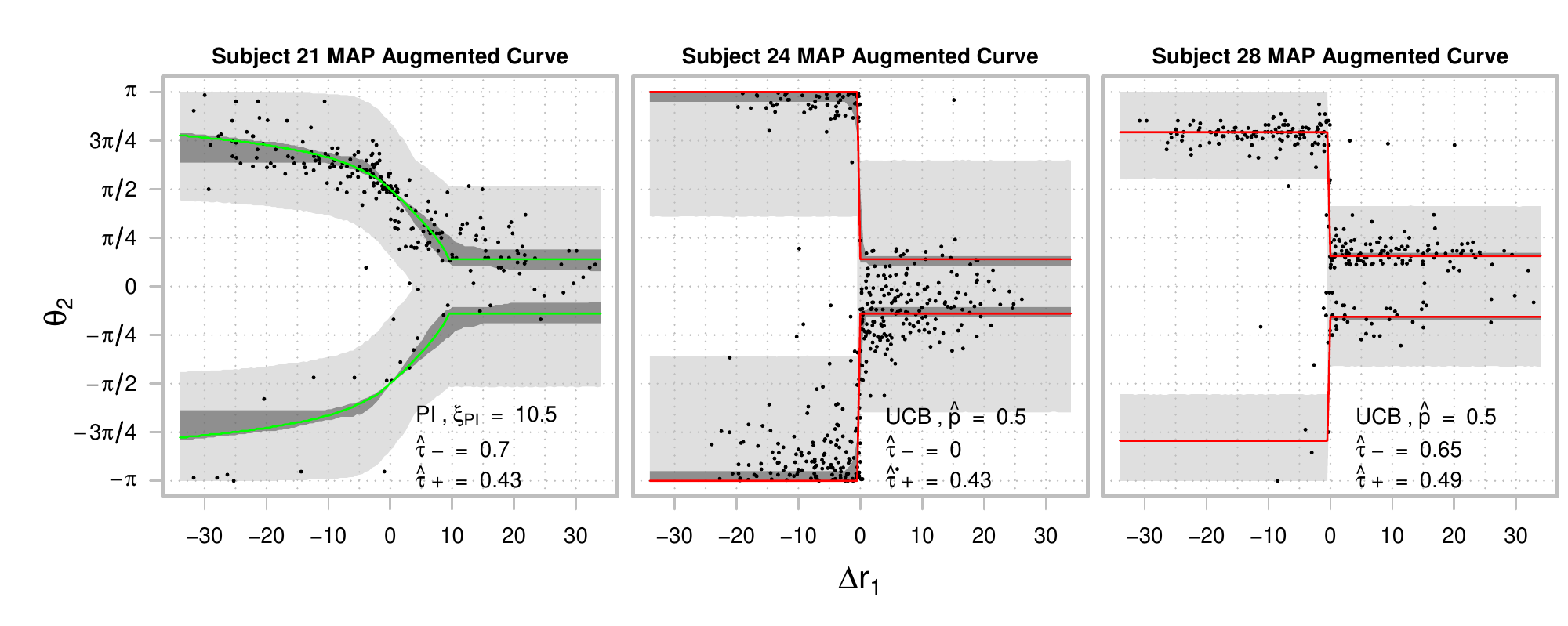}  
\end{center}
\caption{All pairs of ($\Delta r_1$, $\theta_2$) data for subjects 21 (left), 24 (middle), and 28 (right).  Each scatterplot is overlaid with the subject's corresponding MAP augmented acquisition curve in color.  Green denotes PI and red denotes UCB.  Around each curve is an approximate 95\% HPD credible interval on the acquisition curve (dark gray) and 95\% HPD prediction interval (light gray).  Point estimates for the parameters governing each subject's acquisition function are listed in the lower right corner.}
\label{fig:human_acquisition_fits}
\end{figure}

 Figure \ref{fig:human_acquisition_fits} shows the data for the same subjects as before, this time overlaid with the fitted models using the augmented acquisition function in \eqref{eq:threshold_pm}.  Visually, the fit appears to be superior. 

\subsection{Model validation}

We fit the acquisition models in \eqref{eq:optimality_constraint}, \eqref{eq:threshold}, and \eqref{eq:threshold_pm} to each subject's data following the procedure outlined in Section \ref{sec:imperfect_acquisition}.  For the augmented models we estimate the $\tau$ parameters as well. As with the other exploration parameters, we set discrete uniform priors on $\tau$, $\tau^+$, and $\tau^-$.  Table \ref{tab:out_of_sample} shows out-of-sample log-likelihoods (using an 80\%/20\% train/test split of the data) for each model across the subjects in our study.
\begin{table}[h]
\centering
\caption{Out-of-sample log-likelihoods for (\ref{eq:symmetric_distribution}) fit using $u$, $\widetilde{u}$, and $\widetilde{u}^\pm$ for all 28 subjects in our study.  The model(s) with the greatest log-likelihood is shown in bold for each subject.  The bottom row shows the sum of the log-likelihoods across all subjects for each model.}
\medskip
\begin{tabular}{r|ccc||r|ccc}
\toprule
 & \multicolumn{3}{c||} { Acquisition model }  & & \multicolumn{3}{c} { Acquisition model }  \\
 \cline { 2 - 4 }  \cline { 6 - 8 } 
Subject & $u$ & $\widetilde{u}$ & $\widetilde{u}^\pm$ & Subject & $u$ & $\widetilde{u}$ & $\widetilde{u}^\pm$ \\ 
  \hline
1 & \textbf{-72.20} & \textbf{-72.20 }& \textbf{-72.20} &   15 & -71.97 & -70.48 & \textbf{-68.85} \\ 
 2 & -72.84 & -72.95 & \textbf{-67.33} &   16 & -55.97 & -55.97 & \textbf{-35.07} \\ 
  3 & -70.23 & -70.77 & \textbf{-63.85}  &   17 & -52.69 & -47.06 & \textbf{-36.55} \\ 
  4 & -74.76 & -74.76 & \textbf{-70.62} &   18 & -103.88 & -99.60 & \textbf{-78.09} \\ 
  5 & \textbf{-42.42} & \textbf{-42.42} & \textbf{-42.42} &  19 & -139.76 & -138.81 & \textbf{-115.09} \\ 
  6 & -47.16 & 3.37 & \textbf{4.65} & 20 & \textbf{-26.53} & \textbf{-26.53} & -27.06 \\ 
  7 & -73.80 & -68.52 & \textbf{-67.30} &   21 & -50.09 & \textbf{-41.22} & -41.77  \\ 
  8 & -77.99 & -77.99 & \textbf{-70.73}  &   22 & -78.73 & -77.45 & \textbf{-75.40} \\ 
  9 & -34.44 & -34.75 & \textbf{-31.41} &   23 & -122.13 & -122.13 & \textbf{-122.12} \\ 
  10 & -68.14 & -32.20 & \textbf{-21.49} &  24 & -90.92 & -87.96 & \textbf{-81.29} \\ 
  11 & \textbf{-96.68} & \textbf{-96.68} & \textbf{-96.68} & 25 & -36.28 & -24.00 & \textbf{-18.03} \\ 
  12 & -75.46 & -49.71 & \textbf{-47.55} &  26 & \textbf{-80.13} & \textbf{-80.13} & -83.02  \\ 
  13 & -95.25 & -95.25 & \textbf{-95.24} &   27 & \textbf{-69.83} & -74.28 & -75.18 \\ 
  14 & -77.52 & -79.58 & \textbf{-74.29} &   28 & -78.66 & \textbf{-33.51} & -35.21 \\ 
  \hline
  \multicolumn{4}{c} {  }  &   Total & -2036.46 & -1843.54 & \textbf{-1709.19}  \\
\bottomrule
\end{tabular}
\label{tab:out_of_sample}
\end{table}

The $\widetilde{u}^\pm$ model provides the best fit on aggregate, though there are some subjects for which the additional flexibility offered by the $\tau$ parameters does not substantially improve the fit.  Unless otherwise specified, from this point onward all model references utilize the augmented acquisition function in (\ref{eq:threshold_pm}).  
 
We also have included a simulation study in Appendix \ref{sec:simulation_study} of the supplemental material to validate our methodology in a controlled setting.  The study demonstrates that our method accurately recovers the true underlying acquisition function under different acquisition function parameters and observation sample sizes.  It also explores how our method performs when the true acquisition function class is omitted from the inference acquisition function class.

\section{Results} \label{sec:human_tendencies}

Table \ref{tab:results} shows each subject's MAP estimates of the $\widetilde{u}^\pm$ parameters from (\ref{eq:threshold_pm}) in columns 1-3.  Columns 4 and 5 show their estimates of $\gamma$  and $w$, respectively.  Figure \ref{fig:all_fits} in Appendix \ref{sec:additional_figures} of the supplemental material plots the data, MAP augmented acquisition curves, and 95\% HPD intervals for all 28 subjects.
       
One observation that immediately jumps out from the table is that the vast majority of subjects are best represented by the UCB acquisition function with $p = 0.5$.\footnote{The $p = 0.5$ acquisition curve is equivalent to that of $\xi_{\text{\tiny{PI}}} = 0$, hence the posterior mass is equally distributed between these two models.  We chose to report the UCB parameter since this family yielded the best fit for these subjects prior to augmenting the acquisition function.}  Before augmenting the acquisition function, this would imply that most subjects are purely exploitative on their second move, however, after augmenting the acquisition function this interpretation no longer holds.  Depending on a subject's estimated values of $\tau^-$ and $\tau^+$, subjects can exhibit highly exploratory behavior despite having $\hat{p} = 0.5$.  Under the augmented acquisition function, the parameters ${\xi}_{\text{\tiny{PI}}}/{\xi}_{\text{\tiny{EI}}}/{p}$ denote the \textit{shape} of the curve more than the magnitude of their exploration preferences.  Instead, the values of $\tau^-$ and $\tau^+$ primarily explain a subject's exploration vs. exploitation preferences in the augmented formulation.  
\begin{table}[H]
\centering
\caption{Each subject's estimated fit to (\ref{eq:symmetric_distribution}) using the augmented acquisition formulation in \eqref{eq:threshold_pm}.  Columns 1-5 show MAP parameter estimates for the acquisition family shape parameter (${\xi}_{\text{\tiny{PI}}}/{\xi}_{\text{\tiny{EI}}}/{p}$), as well as estimates for $\tau^-$, $\tau^+$, $\gamma$, and $w$.  Columns 6-8 show corresponding measures of subject performance in the search task, where $\overline{\Delta r_i}$ denotes the average change in score after a subject's $i$th move.}
\medskip
\begin{tabular}{r|rrrrr||rrc}
\toprule
 & \multicolumn{5}{c||} { Fitted Model }  & \multicolumn{3}{c} {Task Performance}  \\
 \cline { 2 - 6 }  \cline { 7 - 9 } 
 Subject & $\hat{\xi}_{\text{\tiny{PI}}}/\hat{\xi}_{\text{\tiny{EI}}}/\hat{p}$ & $\hat{\tau^-}$ & $\hat{\tau^+}$ & $\hat{\gamma}$ & $\hat{w}$ & $\overline{\Delta r_2}$ & $\overline{\Delta r_3}$ & $\overline{\sum_{i=2}^{10} \Delta r_i}$ \\ 
  \hline
  1 & $\hat{p}=$ 0.50 & 0.00 & 0.00 & 0.22 & 0.50 & 7.84 & 9.14 & 53.33 \\ 
  2 & $\hat{p}=$ 0.50 & 0.16 & 0.32 & 0.13 & 0.52 & 8.78 & 9.66 & 54.07 \\ 
  3 & $\hat{p}=$ 0.50 & 0.00 & 0.43 & 0.19 & 0.44 & 7.09 & 9.74 & 55.00 \\ 
  4 & $\hat{p}=$ 0.50 & 0.22 & 0.00 & 0.11 & 0.50 & 8.74 & 9.99 & 57.37 \\ 
  5 & $\hat{p}=$ 0.50 & 0.00 & 0.00 & 0.18 & 0.50 & 7.71 & 9.48 & 57.30 \\ 
  6 & $\hat{p}=$ 0.50 & 1.57 & 1.57 & 0.06 & 0.92 & 0.12 & 13.79 & 61.17 \\ 
  7 & $\hat{p}=$ 0.50 & 0.92 & 0.76 & 0.18 & 0.60 & 5.98 & 13.10 & 62.64 \\ 
  8 & $\hat{p}=$ 0.50 & 0.22 & 0.00 & 0.09 & 0.51 & 7.89 & 8.86 & 47.85 \\ 
  9 & $\hat{p}=$ 0.50 & 0.00 & 0.16 & 0.10 & 0.48 & 9.47 & 9.58 & 53.78 \\ 
  10 & $\hat{p}=$ 0.50 & 0.81 & 0.92 & 0.09 & 0.72 & 6.25 & 11.39 & 51.06 \\ 
  11 & $\hat{p}=$ 0.50 & 0.00 & 0.00 & 0.31 & 0.50 & 6.43 & 8.84 & 50.15 \\ 
  12 & $\hat{p}=$ 0.50 & 0.76 & 0.65 & 0.11 & 0.59 & 7.22 & 13.07 & 66.12 \\ 
  13 & $\hat{p}=$ 0.52 & 0.00 & 0.00 & 0.33 & 0.50 & 7.00 & 11.61 & 56.77 \\ 
  14 & $\hat{p}=$ 0.50 & 0.22 & 0.76 & 0.23 & 0.55 & 5.54 & 10.11 & 49.29 \\ 
  15 & $\hat{p}=$ 0.50 & 0.32 & 0.60 & 0.29 & 0.45 & 7.07 & 8.11 & 46.09 \\ 
  16 & $\hat{p}=$ 0.50 & 0.00 & 0.60 & 0.13 & 0.36 & 9.98 & 13.40 & 72.27 \\ 
  17 & $\hat{p}=$ 0.50 & 0.11 & 0.43 & 0.11 & 0.50  & 8.95 & 9.38 & 59.03 \\ 
  18 & $\hat{p}=$ 0.50 & 1.57 & 0.54 & 0.27 & 0.46 & 3.63 & 10.22 & 50.49 \\ 
  19 & $\hat{p}=$ 0.50 & 1.30 & 0.70 & 0.22 & 0.45 & 4.09 & 9.53 & 47.42 \\ 
  20 & $\hat{p}=$ 0.50 &  0.16 & 0.00 & 0.14 & 0.50 & 9.13 & 10.53 & 56.24 \\ 
  21 & $\hat{\xi}_{\text{\tiny{PI}}} =$ 10.5 & 0.70 & 0.43 & 0.15 & 0.87 & 6.19 & 15.16 & 66.54 \\ 
  22 & $\hat{p}=$ 0.50 & 0.22 & 0.27 & 0.19 & 0.53 & 8.37 & 10.74 &  61.12 \\ 
  23 & $\hat{p}=$ 0.50 &  0.00 & 0.00 & 0.37 & 0.50 & 7.01 & 9.49 & 48.60  \\ 
  24 & $\hat{p}=$ 0.50 & 0.00 & 0.43 & 0.24 & 0.41 & 8.83 & 10.95 & 59.73 \\ 
  25 & $\hat{p}=$ 0.50 & 0.81 & 0.97 & 0.07 & 0.10 & 6.93 & 10.65 & 50.55  \\ 
  26 & $\hat{p}=$ 0.50 & 0.22 & 0.00 & 0.18 & 0.48& 8.44 & 9.66 & 58.67 \\ 
  27 & $\hat{\xi}_{\text{\tiny{PI}}} =$ 7 & 0.97 & 0.87 & 0.29 & 0.32 & 4.06 & 12.31 & 55.49  \\ 
  28 & $\hat{p}=$ 0.50 &0.65 & 0.49 & 0.10 & 0.83 & 6.99 & 14.37 & 70.59 \\ 
\bottomrule
\end{tabular}
\label{tab:results}
\end{table}     
The estimated values of $\tau^-$ and $\tau^+$ in Table \ref{tab:results} show that while many subjects exhibit strong exploratory preferences, there is not a universal pattern exhibited across all of the subjects.  Many subjects appear highly exploratory regardless of the change in score on move 1 (subjects 6, 7, 10, 12, 25, and 27) while even more subjects are always exploitative (subjects 1, 4, 8, 9, 11, 13, 20, 23, and 26).   Some subjects had significantly stronger exploratory preferences when their first move improved their score (subjects 14 and 16) while others were more likely to explore when their first move yielded a negative change in their score (subjects 18 and 19).  

Another behavioral tendency we can assess is the subjects' preferences toward either turning left or right on move 2 (relative to the direction of their first move).  Based on the estimates of $w$ shown in Table \ref{tab:results}, we find that most subjects do not exhibit a strong bias for one direction over the other, but some subjects do have a strong preference to turn right after move 1 (subjects 6, 10, 21, 28) and one subject has a strong preference to turn left (subject 25).      

While the focus of our paper is on making inference on human acquisition strategies when facing an exploration vs exploitation conflict, there are a few relationships between the estimated acquisition functions and subjects' corresponding performance in the task which we will briefly discuss.  First, there is a strong relationship between high exploitation preferences and success in the early moves of the task.  Letting $\widehat{\overline{\tau}}_j = (\widehat{\tau}^-_j + \widehat{\tau}^+_j)/2$ and $\overline{\Delta r_2}_j$ denote the average change in score after the second move across all of subject $j$'s rounds of the task, the correlation between $ \widehat{\overline{\boldsymbol \tau}}$ and $\overline{\Delta \boldsymbol r_2}$ is -0.81.  This is not surprising---subjects who explore on their second move naturally get a lower average score on this move. However, this exploratory sacrifice is balanced by the ability to better exploit the resulting information on the remaining moves in the round, as reflected in the correlation($\widehat{\overline{\boldsymbol \tau}}$, $\overline{\Delta \boldsymbol r_3}$) = 0.49.  Though the relationship is not equally strong, high explorers perform better on average on moves 3 and up.  

Ultimately, we care most about the total score improvement on all informed moves: $\sum_{i=2}^{10} \Delta r_i$.  We do not find a significant relationship between the estimated acquisition function parameters and this measure.  However, a strong relationship exists between $\sum_{i=2}^{10} \Delta r_i$ and $\sum_{i=4}^{10} |\Delta \theta_i|$, where $\Delta \theta_i = \theta_i - \theta_{i-1}$. This measures how much ``zig-zag" a subject exhibits after having the necessary information to estimate a planar gradient over the click-region.  As explained in Appendix \ref{sec:move_3} of the supplemental material, given a linear model for the surrogate, the Bayesian optimization algorithm will no longer yield drastic changes in direction after the 3rd move.  By contrast, most subjects showed significantly positive values of $\sum_{i=4}^{10} |\Delta \theta_i|$.  This could indicate preferences for continued exploration, but it could also result from other unintentional sources of sampling variation (e.g. sloppiness due to the speed with which subjects performed the task).  Not surprisingly, the degree of ``zig-zag" is related to the aggregate performance measure: correlation($\sum_{i=4}^{10} |\Delta \theta_i|$, $\sum_{i=2}^{10} \Delta r_i$) = -0.47.  Subjects who kept a more constant direction after their 3rd move performed better in the task.  

\section{Conclusion} \label{sec:conc}

This work introduces a probabilistic framework for inverse Bayesian optimization and shows how it can be implemented through a lab-based sequential optimization experiment.  Our methodology involves defining a likelihood on a subject's observed behavior from a search task, where the likelihood is parameterized by a Bayesian optimization subroutine governed by an acquisition function with parameters of its own.  This novel structure enables us to make inference on the subject's acquisition function while allowing their behavior to deviate around the solution to the underlying Bayesian optimization subroutine.  This allows us to simultaneously make probabilistic inference about the subject's acquisition function in addition to providing a predictive model of their behavior.  The combination of an optimization subroutine within a likelihood-based predictive model distinguishes our approach from other inverse optimization methods.  

There are numerous directions that could be explored in future work.  In this paper, we base our inference on the 2nd acquisition conditional on the first, but in theory all of a subject's decisions (acquisitions) in a given sample path could be leveraged when making this inference.  How to incorporate all of this information into the inference on the acquisition function is a promising area of further study.  In addition, our work focuses on a highly customized Bayesian optimization framework.  A more general treatment of the inverse Bayesian optimization problem (including GP surrogates, higher dimensionality, a richer class of acquisition functions, etc.) represents a rich and challenging area of future work.

Finally, our study supports findings in human cognition.  Specifically, we find that subjects exhibit a wide array of acquisition preferences, but that nearly all of them exhibit exploration tendencies beyond the ability of standard acquisition functions to capture.  As in \cite{wu2018generalization}, additional exploratory modifications must be included in the acquisition model in order to accurately represent the observed human behavior.  Consistent with \cite{borji2013bayesian}, \cite{wu2018generalization}, and \cite{candelieri2020modelling}, we find that UCB acquisition functions best represent the search strategies of the majority of subjects in our study.    
 
\bibliographystyle{ba}
\bibliography{bibliography}

\section*{Acknowledgements}
We would like to thank Derek Bingham, Nasrin Yousefi, the associate editor, and two anonymous reviewers for helpful comments and suggestions on the paper. 

\begin{appendix}

\section{Moves 3 and up} \label{sec:move_3}
 
Once a subject has made their second move they have a near-complete characterization of the objective function in the local region defined by the move 3 click-region.  After receiving feedback from move 2, subjects can approximate the objective function locally with high fidelity using the plane defined by moves 0, 1, and 2.  In theory, as long as their second move is not along the same direction as move 1, they can solve for the direction of the hotspot.  

\begin{figure}[H]
\begin{center}
\includegraphics[trim={0cm .5cm 0cm 0cm}, clip, width=1\textwidth]{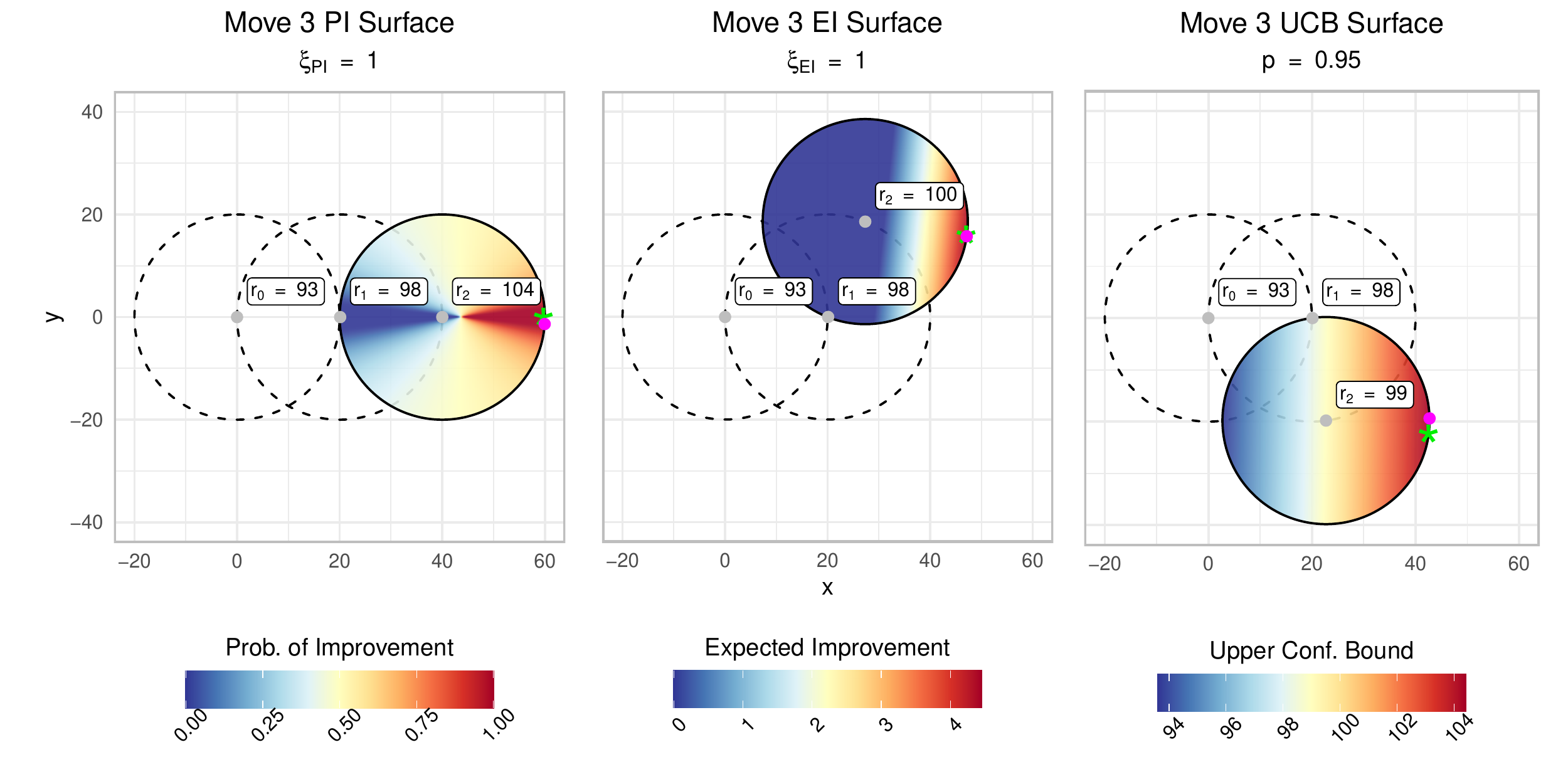} 
\end{center}
\caption{Move 3 acquisition surfaces for three sample acquisition functions.  In each plot, $r_0 = 93$ and $r_1 = 98$, but the $r_2$ values are different because each $\mathbf{m}_2$ was selected at an optimum of the previous iteration of the Bayesian optimization process (i.e. the green stars in Figure \ref{fig:example_acquisition_m2} of the main text).  The $\argmax$ of each surface is denoted by a green star while the pink dots show the true direction of the hotspot.}
\label{fig:example_acquisition_m3}
\end{figure}

This phenomenon is illustrated in Figure \ref{fig:example_acquisition_m3}.  These plots are the same as those in Figure 4 of the main text, only here we plot the move 3 acquisition surfaces instead of the move 2 surfaces.  In each case, move 2 was selected at an optimum of the previous iteration of the process.  The green stars denote the $\argmax$ of each surface, while the pink dots show the true direction of the hotspot.  In every case, the green stars are nearly identical to the pink dots.   

\begin{figure}[H]
\begin{center}
\includegraphics[trim={0cm 1.25cm 0cm .5cm}, clip, width=1\textwidth]{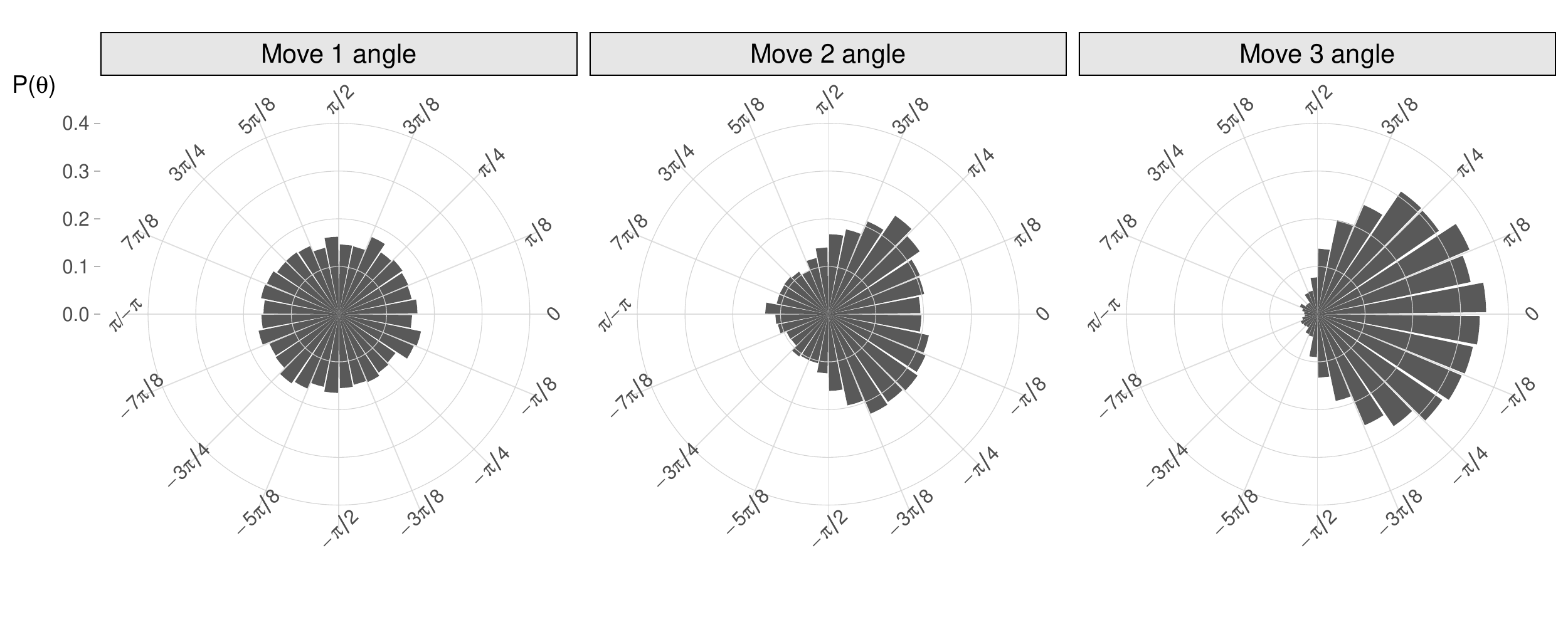} 
\end{center}
\caption{Histograms of subjects' move angles relative to hotspot direction on moves 1, 2, and 3, aggregated over all subjects and rounds.}
\label{fig:angle_to_hotspot}
\end{figure}

While it is possible to learn the direction of the hotspot after move 2,  whether subjects actually do learn the optimal direction is a separate question.  Figure \ref{fig:angle_to_hotspot} shows directional histograms of subject behavior on moves 1, 2, and 3 relative to the direction of the hotspot (aggregated over all subjects).  On move 1, the angles are uniform on the circle because subjects have no directional information about the hotspot on this move.  The move 2 angles are directionally informed, but subjects must move off the axis of the informed direction in order to gain information about the orthogonal gradient.   This can be seen in the move 2 histogram by the modes over $\pm \pi/4$.   

The move 3 histogram sheds light on how well subjects synthesize the information from moves 1 and 2 in order to learn the direction of the hotspot.  As shown by the variance around 0 in the move 3 histogram, subjects do not always synthesize this information perfectly, but the mode angle is toward the direction of the hotspot.  On moves 4 through 10 (not shown), the mode remains centered around the direction of the hotspot while the variance around the mode decreases slightly on each subsequent move.  Since subject behavior  essentially amounts to a noisy walk around the optimal direction after move 2,  moves 3 and up provide comparatively little information about the subjects' acquisition preferences.  For this reason we do not analyze these moves in our paper. 

\section{IBO prior specification} \label{sec:prior_specification}

We use discrete uniform priors on the parameters defining the likelihood in (4.4)-(4.6).  In the notation that follows, $\mathcal{U}_n\{a,b\}$ denotes a discrete uniform distribution spanning $a$ to $b$ over $n$ equally spaced intervals. 
\begin{align}
\xi_{\text{\tiny{PI}}}, \xi_{\text{\tiny{EI}}}  &\sim \mathcal{U}_{61}\{0, 30\} \\
p &\sim \mathcal{U}_{61}\{0.5, 0.99\} \\
\gamma &\sim \mathcal{U}_{61}\{0.01, \pi/4\} \\
\tau, \tau^-, \tau^+ &\sim \mathcal{U}_{61}\{0, \pi/2]\} 
\end{align}

\section{Simulation Study} \label{sec:simulation_study}

We performed a simulation study to explore how well our method can recover a latent acquisition function used to generate $\theta_2$ (move 2 angles) conditional on randomly generated values of $\Delta r_1$ in a number of settings that include different acquisition function parameters and observation sample sizes.  We also explored how our method performs when the true generating acquisition function class is omitted from the inference acquisition function class.  

\subsection{Data generation}

In each simulation, we generated ($\Delta r_1^i$, $\theta_2^i$) pairs for $i = 1,\ldots,n_{obs}$.  These pairs represent move 2 acquisitions conditional on move 1 scores over $n_{obs}$ rounds of the task, where $n_{obs} \in \{10, 100, 1000\}$.  We generated these pairs as follows for a given acquisition function $u$:
\begin{enumerate}
    \item Draw $\Delta r_1^i \sim Unif(-34, 34)$.  This is the approximate range of $\Delta r_1$ in the hotspot experiment.  
    \item Compute $\hat{f}_{1}$ via \eqref{eq:post_beta}-\eqref{eq:post_pred} % 
    \item Compute $\argmax \limits_{\theta \in (-\pi, \pi]} u(\theta | \hat{f}_1, \Delta r_1^i)$
    \item Draw $\theta_2^i \sim \ell(\theta_{2} \mid \Delta r_{1}, u, \hat{f}_{1}, \gamma, w)$ via \eqref{eq:symmetric_distribution}, where
    \begin{itemize}
        \item $w = 0.5$
        \item $\gamma$ = 0.25
    \end{itemize}
    These parameters were not observed to affect the coverage values of the acquisition function parameters, thus we set them arbitrarily at reasonable values within the range of the parameter estimates for the subjects in our study.   
\end{enumerate}
We generated these pairs for the PI, EI, and UCB acquisition functions with the following exploration parameters:
    \begin{itemize}
        \item $\xi_{PI} \in \{1, 15, 30\}$
        \item $\xi_{EI} \in \{0, 15, 30\}$
        \item $p \in \{0.5, 0.75, 0.99\}$
    \end{itemize}
    
\subsection{Inference class contains the truth}

After generating the data for each acquisition function/$n_{obs}$ combination, we fit the data to the classes of acquisition functions defined in \eqref{eq:PI}-\eqref{eq:UCB} following the procedure outlined in Section \ref{sec:imperfect_acquisition} of the main text.  Table \ref{tab:truth} shows the acquisition function coverage values based on 95\% credible regions of the corresponding posterior distribution over 1000 simulations.  Each cell represents the proportion of 95\% credible acquisition functions regions that covered the true function that generated the data.
\begin{table}[H]
\centering
        \begin{tabular}{cl|ccc}
        \toprule
        \multirow{2}{*}{Acquisition}  & \multirow{2}{*}{$\xi / p$ value} &  \multirow{2}{*}{$n_{obs}$ = 10} & \multirow{2}{*}{$n_{obs}$ = 100} & \multirow{2}{*}{$n_{obs}$ = 1000} \\
       & \\
        \hline
        \hline
            \multirow{3}{*}{PI} & $\xi_{PI} = 1$ & 0.94 & 0.93 & 0.92 \\
                                 & $\xi_{PI} = 15$ & 0.97 & 0.96 & 0.85 \\
                                 & $\xi_{PI} = 30$ & 0.99 & 0.96 & 0.98 \\

        \hline
            \multirow{3}{*}{EI} & $\xi_{EI} = 0$ & 0.97 & 0.94 & 0.92 \\
                                 & $\xi_{EI} = 15$ & 0.97 & 0.93 & 0.94 \\
                                 & $\xi_{EI} = 30$ & 0.99 & 0.99 & 0.99 \\
        \hline
            \multirow{3}{*}{UCB} & $p = 0.5$ & 0.89 & 0.83 & 0.99 \\
                                 & $p = 0.75$ & 0.98 & 0.98 & 0.93 \\
                                 & $p = 0.99$ & 0.99 & 0.99 & 0.99 \\
   \bottomrule
        \end{tabular}
    \caption{Acquisition function coverage values based on 95\% credible regions constructed via the inversion method described in Section \ref{sec:imperfect_acquisition} for various acquisition function (rows) and number of observations (columns) combinations.}
    \label{tab:truth}
\end{table}
As shown in the table, the coverage values tend to hover around 95\% across the simulations, showing that our method accurately recovers the true generating acquisition function.  

\subsection{Inference class omits the truth}

We next explored how our method performs when the true generating acquisition function class is omitted from the inference acquisition function class.  For this study, we likewise fit the data to the classes of acquisition functions defined in \eqref{eq:PI}-\eqref{eq:UCB} following the procedure outlined in Section \ref{sec:imperfect_acquisition}, only we omitted the acquisition family that was actually used to generate the data when performing the inference.  Table \ref{tab:omit} shows the proportion of 95\% credible regions ($CR$) from the resulting inference that contain parameter values from the two acquisition families that were not used to generate the data (over 1000 simulations).  Note that the two coverage values on the same row of a given cell do not have to sum to 1.  

\begin{table}[H]
\centering
        \begin{tabular}{cl|ll|ll|ll}
        \toprule
        \multirow{2}{*}{Acq.}  & \multirow{2}{*}{$\xi / p$ value} &  \multicolumn{2}{c}{\multirow{2}{*}{$n_{obs}$ = 10}} & \multicolumn{2}{c}{\multirow{2}{*}{$n_{obs}$ = 100}} & \multicolumn{2}{c}{\multirow{2}{*}{$n_{obs}$ = 1000}} \\ 
        & \\
        \hline
        \hline
                                   &   & \scriptsize{$\xi_{EI} \in CR$} & \scriptsize{$p \in CR$} & \scriptsize{$\xi_{EI} \in CR$} & \scriptsize{$p \in CR$} & \scriptsize{$\xi_{EI} \in CR$} & \scriptsize{$p \in CR$} \\ \cline{3-8}
            \multirow{3}{*}{PI} & $\xi_{PI} = 1$ & 0.47 & 1.00 & 0.00 & 1.00 & 0.00 & 1.00 \\
                                 & $\xi_{PI} = 15$ & 0.99 & 0.83 & 1.00 & 0.00 & 1.00 & 0.00 \\
                                 & $\xi_{PI} = 30$ & 1.00 & 0.93 & 1.00 & 0.03 & 1.00 & 0.00 \\
        \hline
                                           &   & \scriptsize{$\xi_{PI} \in CR$} & \scriptsize{$p \in CR$} & \scriptsize{$\xi_{PI} \in CR$} & \scriptsize{$p \in CR$} & \scriptsize{$\xi_{PI} \in CR$} & \scriptsize{$p \in CR$} \\\cline{3-8}
            \multirow{3}{*}{EI} & $\xi_{EI} = 0$ & 0.99 & 0.84 & 1.00 & 0.01 & 1.00 & 0.00\\
                                 & $\xi_{EI} = 15$& 0.99 & 0.88 & 1.00 & 0.01 & 1.00 & 0.00 \\
                                 & $\xi_{EI} = 30$ & 0.96 & 0.99 & 0.39 & 0.90 & 0.16 & 0.90 \\
        \hline
                                           &   & \scriptsize{$\xi_{PI} \in CR$} & \scriptsize{$\xi_{EI} \in CR$} & \scriptsize{$\xi_{PI} \in CR$} & \scriptsize{$\xi_{EI} \in CR$} & \scriptsize{$\xi_{PI} \in CR$} & \scriptsize{$\xi_{EI} \in CR$} \\\cline{3-8}
            \multirow{3}{*}{UCB} & $p = 0.5$ & 1.00 & 0.40 & 1.00 & 0.00 & 1.00 & 0.00\\
                                 & $p = 0.75$ & 1.00 & 1.00 & 0.98 & 0.97 & 0.01 & 1.00\\
                                 & $p = 0.99$ & 0.99 & 1.00 & 0.01 & 1.00 & 0.00 & 1.00\\
   \bottomrule
        \end{tabular}
    \caption{Proportion of 95\% credible regions ($CR$) that contain parameter values from the acquisition families that were not used to generate the data (over 1000 simulations).  The acquisition family used to generate the data is shown in the left hand column and the number of observations used in the inference is shown in the top row.  For each simulation, the acquisition family used to generate the data was not included in the inference class.}
    \label{tab:omit}
\end{table}

When the inference is based on a small amount of data ($n_{obs} = 10$) the resulting credible regions tend to contain acquistion functions in both of the misspecified classes.  As the number of observations increases, the credible regions tend to converge to a single acquisition family, which presumably is closer to the true shape of the omitted generating function.  No single family appears to dominate the competing family for every simulation setting. 

\section{GitHub repository}

All of our data and code to reproduce our results are hosted publicly at the first author's GitHub page: \url{https://github.com/nsandholtz/hotspot_paper}.

\section{Additional figures} \label{sec:additional_figures}

\begin{figure}[H] 
\begin{center}
\includegraphics[trim={0cm 4.7cm 0cm 4.5cm}, clip, width=.325\textwidth]{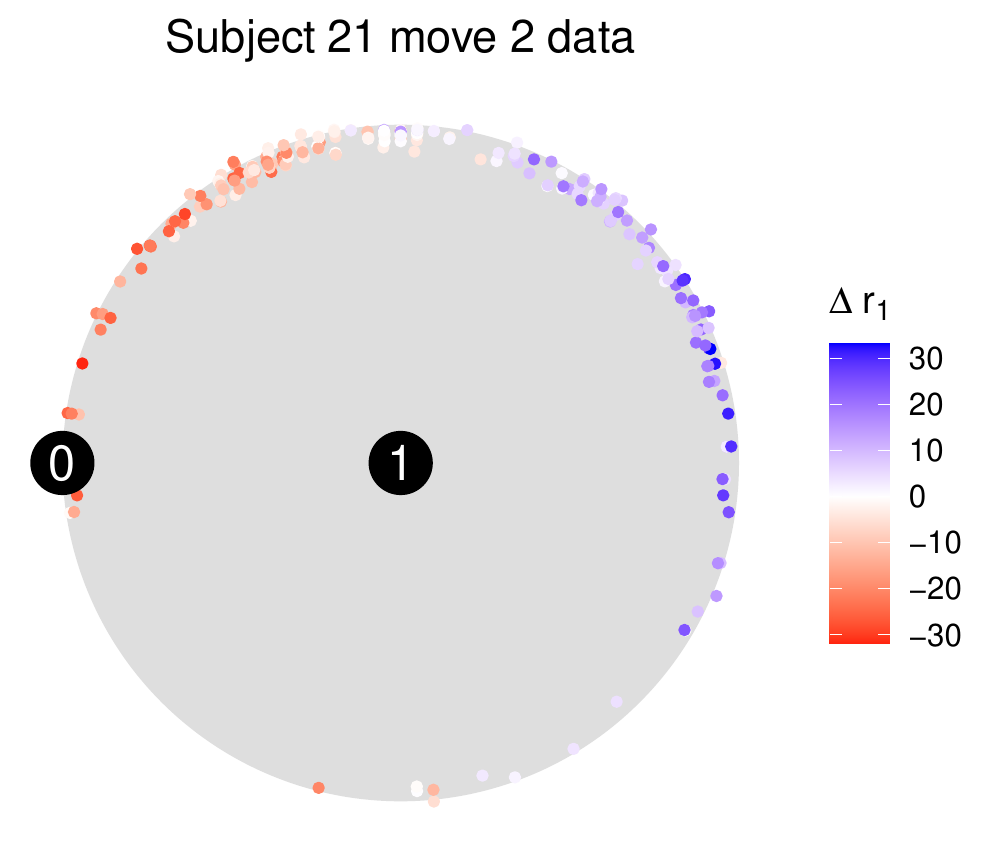} 
\includegraphics[trim={0cm 4.7cm 0cm 4.5cm}, clip, width=.325\textwidth]{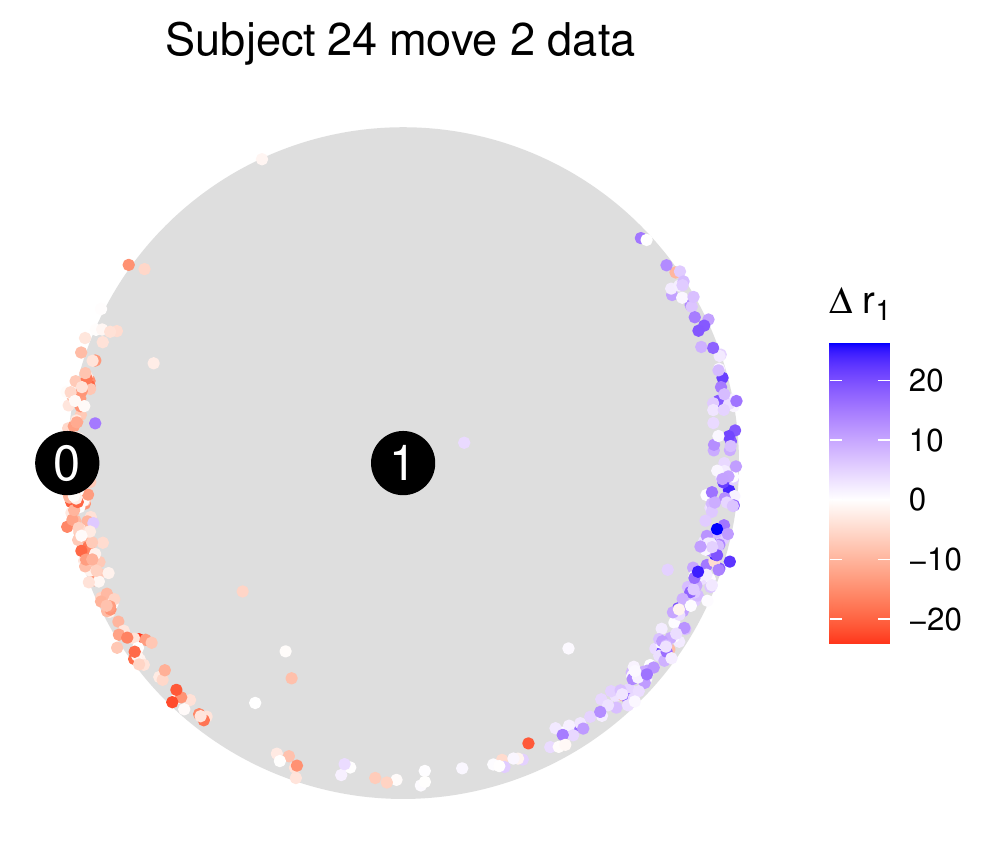}
\includegraphics[trim={0cm 4.7cm 0cm 4.5cm}, clip, width=.325\textwidth]{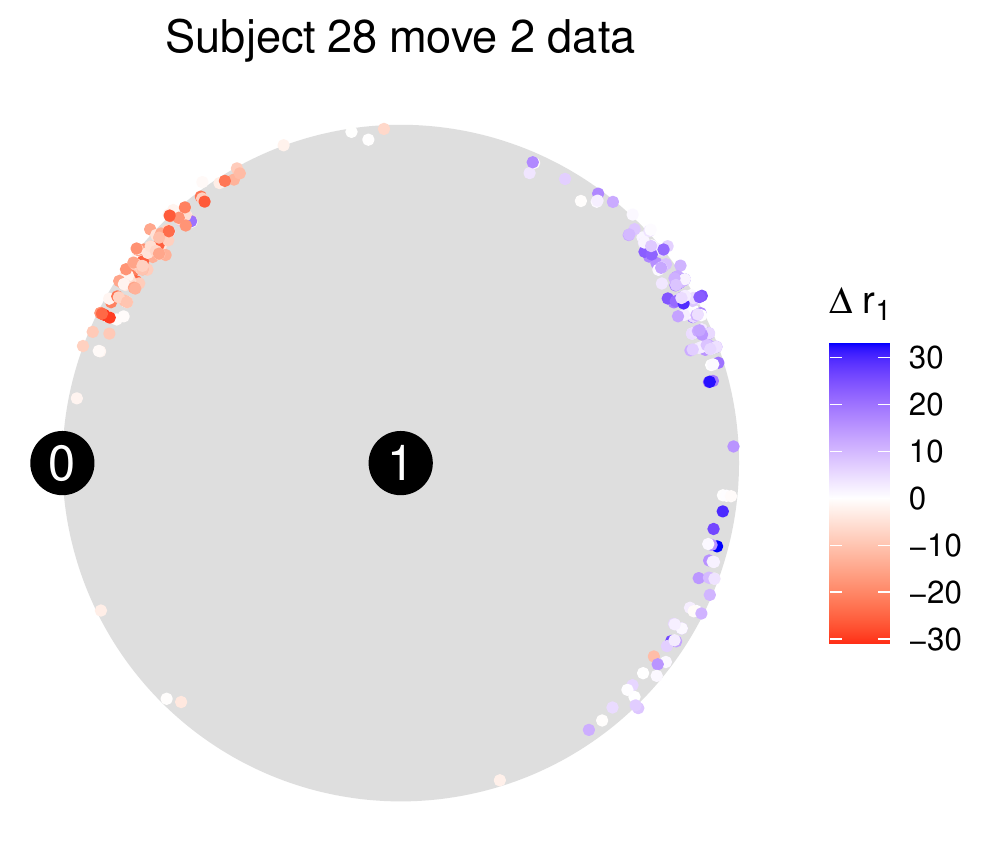}
\end{center}
\caption{Move 2 ($x$, $y$) pairs for three subjects in the experiment.  On each plot, the large black dots labeled 0 and 1 show the starting location and first move location, respectively (the first move is shown along the $x$-axis in order to illustrate trends in the move 2 behavior).  The smaller colored dots show the locations of the subjects' second moves relative to their first moves.  The color gradient denotes $\Delta r_1$.}
\label{fig:move_2_xy}
\end{figure} 

\begin{figure}[H]
\begin{center}
\includegraphics[trim={.55cm .75cm .05cm .05cm}, clip, width=1\textwidth]{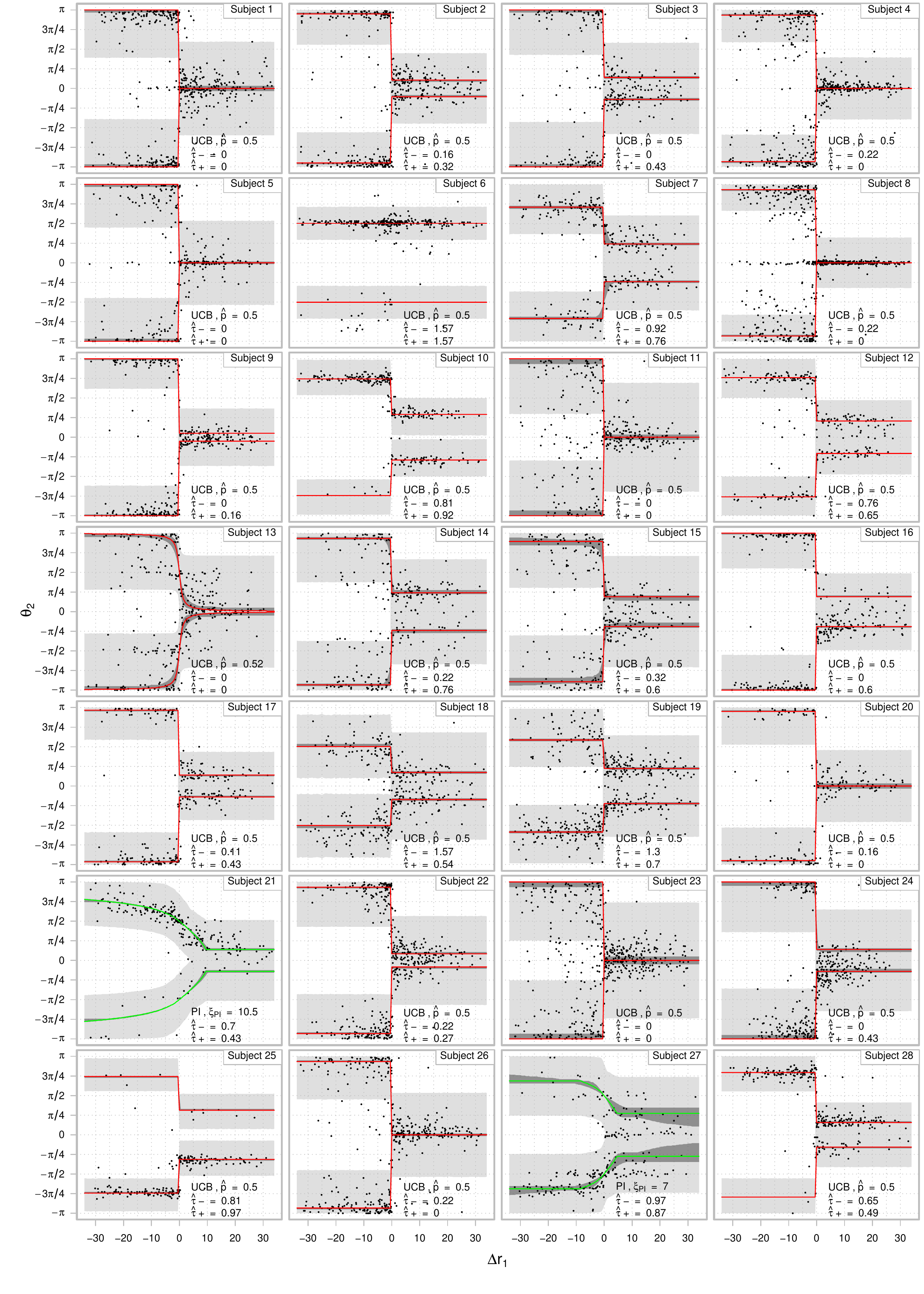}
\end{center}
\caption{($\Delta r_1$, $\theta_2$) pairs for each subject in the study, overlaid with the subject's MAP augmented acquisition curve, 95\% HPD credible regions of the acquisition curve (dark-gray), and 95\% HPD prediction regions (light-gray).  Green denotes PI and red denotes UCB.}
\label{fig:all_fits}
\end{figure}

\end{appendix}

\end{document}